\documentclass[reprint,aps,prb,superscriptaddress]{revtex4-1}
\usepackage{amsmath}
\usepackage{amssymb}
\usepackage{natbib}
\usepackage{graphicx}
\usepackage{xcolor}
\usepackage{amsfonts}

\setcitestyle{numbers,square}

\begin{document}

\title{Optimal mode matching in cavity optomagnonics} \author{Sanchar Sharma} \affiliation{Kavli Institute of NanoScience, Delft University of Technology, 2628 CJ Delft, The Netherlands}  

\author{Babak Zare Rameshti} \affiliation{Department of Physics, Iran University of Science and Technology, Narmak, Tehran 16844, Iran}  

\author{Yaroslav M. Blanter} \affiliation{Kavli Institute of NanoScience, Delft University of Technology, 2628 CJ Delft, The Netherlands}  

\author{Gerrit E. W. Bauer} \affiliation{Institute for Materials Research \& WPI-AIMR \& CSRN, Tohoku University, Sendai 980-8577, Japan} \affiliation{Kavli Institute of NanoScience, Delft University of Technology, 2628 CJ Delft, The Netherlands}  

\begin{abstract}
Inelastic scattering of photons is a promising technique to manipulate magnons but it suffers from weak intrinsic coupling. We theoretically discuss an idea to increase optomagnonic coupling in optical whispering gallery mode cavities, by generalizing previous analysis to include the exchange interaction. We predict that the optomagnonic coupling constant to surface magnons in yttrium iron garnet (YIG) spheres with radius $300\,\mathrm{\mu}$m can be up to $40$ times larger than that to the macrospin Kittel mode. Whereas this enhancement falls short of the requirements for magnon manipulation in YIG, nanostructuring and/or materials with larger magneto-optical constants can bridge this gap.

\end{abstract}

\maketitle

Magnetic insulators such as yttrium iron garnet (YIG) are promising for future spintronic applications such as low power logic devices \cite{Chumak15}, long-range information transfer \cite{Cornelissen15}, and quantum information \cite{Tabuchi16}. Their excellent magnetic quality \cite{Cherepanov_YIG,WuHoff} implies spin waves or magnons, the excitations of the magnetic order, are long-lived. Microwaves in high quality cavities and striplines couple strongly to magnons with long (mm) wavelengths \cite{SoykalPRL10,HueblPol,TabuchiHybrid14,Zhang14,YunshanPol,BabakPol,Babak18}, i.e. the rate of energy exchange between the two systems is higher than their individual dissipation rates, but not to short wavelengths (except under special geometries \cite{JamesMW}). Magnons can be injected electrically by metallic contacts \cite{Kajiwara10,SMR_NL1}, but only in rather small numbers. Here, we focus on the coherent coupling of magnetic order and infrared laser light with sub-$\mathrm{\mu}$m wavelengths, that is enhanced by using the magnet as an optical cavity \cite{ZhangWGM16,Osada16,JamesWGM}. 

By the high dielectric constant and almost perfect transparency in the infrared \cite{WoodRemeika,Lacklison}, sub-mm YIG spheres support long-living whispering gallery modes (WGMs) \cite{James15,ZhangWGM16}. The photons, with energy deep within the band gap, scatter inelastically by absorbing or creating magnons \cite{Wettling75,Borovik82}. This is known as Brillouin light scattering (BLS) \cite{Sebastian15Rev}, which is enhanced in an optical cavity \cite{James15,ZhangWGM16,Osada16,JamesWGM,Hisatomi16,Osada18_Exp,JamesLowL,Pantazopoulos,MagPhotonRev}. These results led to predictions of the Purcell effect \cite{Tianyu16} (optically induced enhancement of magnon linewidth), magnon lasing \cite{Silvia16} and magnon cooling \cite{OMagCool}. However, the models addressed only the magnetostatic magnon modes, i.e. ignored retardation and the exchange interaction, with only small overlap with the WGMs \cite{ZhangWGM16,Osada16,JamesWGM,Hisatomi16,WGMOptoMag,Osada18_Th,MagPhotonRev}. Thus, the observed and predicted coupling rates were too low to be able to optically manipulate magnons \cite{Silvia16,OMagCool}. Higher optomagnonic coupling can be achieved by reducing the size of the magnets down to optical wavelengths \cite{Evangelos}, but this requires nanostructuring of the magnet \cite{Berzhansky13,Chang14,Hauser16}. Coupling to magnons in a non-uniform magnetization texture is large \cite{Graf_Vortex}. Here, we suggest and analyze a method to increase coupling in a conventional set-up of a uniformly magnetized sub-mm YIG sphere by coupling to exchange-dipolar modes with wavelengths comparable to the WGMs. 

Bulk magnons in films with both exchange and dipolar interactions have been extensively studied \cite{StanPrabh,KalinikosSlavin,CylinderMagnons18}. In thick films, exchange reduces the life time of surface magnons by mixing with bulk states \cite{WamesWolfram_Linewidth,CamleyMills,CamilloPaolo}, while in thinner films exchange leads to modes with partial bulk and surface character \cite{WamesWolfram_Detailed,CamilloPaolo}. Here, we address magnetic spheres with radii that are large enough to support surface exchange-dipolar magnons.  

\begin{figure}[ptb]
\begin{equation*}
\includegraphics[width=\columnwidth]{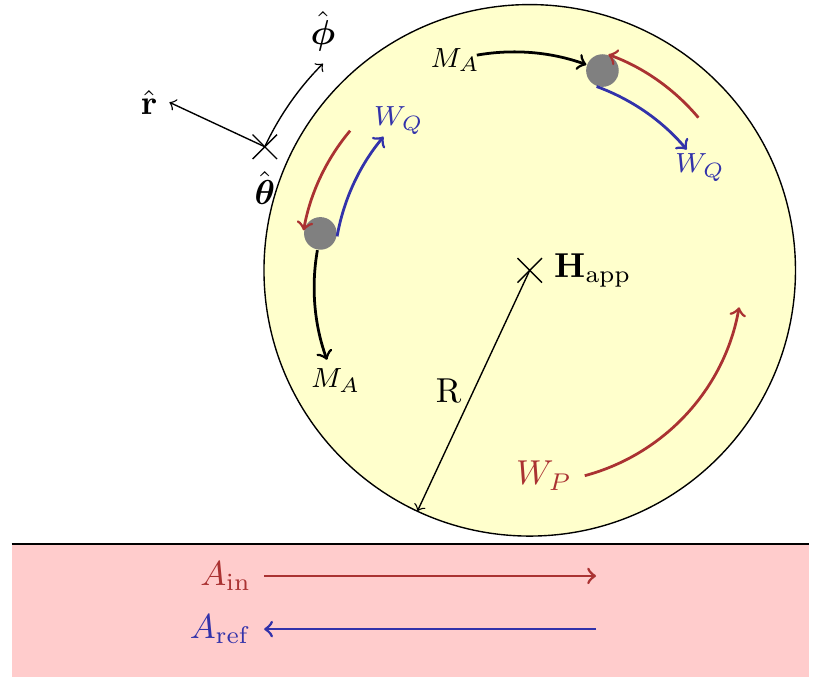}
\end{equation*}
\caption{A sphere made of a ferromagnetic dielectric in proximity to an optical fiber or prism. A magnetic field saturates the magnetization. The input photons in the fiber, $A_{\mathrm{in}}$, leak into the whispering gallery modes (WGMs) $\{W_{P}\}$. The latter can be reflected by magnons $\{M_{A}\}$ of twice the angular momentum into the blue, via $W_{P}+M_{A}\rightarrow W_{Q}$, or red, $W_{P}\rightarrow W_{Q}+M_{A}$, sideband. The photons $\{W_{Q}\}$ can leak back into the fiber and be observed in the reflection spectrum. }
\label{Fig:Sys}
\end{figure}

Our system is sketched in Fig. \ref{Fig:Sys}. A ferromagnetic sphere acts as a WGM resonator in which photons interact with the magnetic order via standard proximity coupling to an optical prism or fiber. The frequency of photons is 4 to 5 orders of magnitude larger than magnons at similar wavelengths, thus the incident and scattered photons have nearly the same frequency and wavelength. \emph{Forward scattering} of photons occurs via magnons of large wavelength $\sim100\,\mathrm{\mu}$m, which is a process that is well described by a purely dipolar theory \cite{WGMOptoMag}. Here we discuss \emph{back scattering} of photons by magnons with sub$\mathrm{- \mu}$m wavelengths that are affected significantly by exchange. We show that the exchange generates magnetic modes that have a near ideal overlap with the optical WGMs, with an optomagnonic coupling limited only by the bulk magneto-optical constants.  

We first briefly review the basics of cavity optomagnonics and derive an upper bound for the optomagnonic coupling constant in resonators in Sec. \ref{Sec:OMag}. We model the magnetization dynamics by the Landau-Lifshitz equation introduced in Sec. \ref{Sec:LL}. The spatial amplitude of surface exchange-dipolar magnons is discussed in Sec. \ref{Sec:SolDisc}, with details of the derivation in App. \ref{App:SolvingMagnons}. The optomagnonic coupling constants found in Sec. \ref{Sec:Coup} are compared with the upper bound found in Sec. \ref{Sec:OMag}. We conclude with discussion and outlook in Sec. \ref{Sec:Disc}.  

\section{Cavity optomagnonics} \label{Sec:OMag}

Here we summarize the basic theory of magnon-photon coupling in spherical optical resonators\ \cite{WGMOptoMag}. The electric and magnetic fields of the optical modes in a spherical resonator are labeled by orbital indices $\{l,m,\nu\}$ and a polarization $\sigma\in\{\text{TM},\text{TE}\}$. They become optical whispering gallery modes (WGMs) at extremal cross sections when $l,m\gg\left\{1,|l - m|\right\}$. WGMs are traveling waves in the $\pm\phi$-direction with dimensionless wavelength $2\pi/m . $ $\nu - 1$ and $l - m $ are the number of nodes in the optical fields in the $r$ and $\theta$ direction. The electric field of these modes is $\mathbf{E}_{\mathrm{TM}} = E(\mathbf{r})\hat{\boldsymbol{\theta}}$ and $\mathbf{E}_{\mathrm{TE}} = E(\mathbf{r})\hat{\mathbf{r}}$ where \cite{Oraevsky_WGM}, 
\begin{equation}
	 E(\mathbf{r}) = \mathcal{E}Y_l^m(\theta,\phi)J_l(kr) . \label{WGM:Elec} 
\end{equation} 
Here $J_l$ is the Bessel function of order $l$ [Eq.~(\ref{Def:Bess})] and $Y_l^m$ is a scalar spherical harmonic [Eq.~(\ref{Def:Ylm})]. The wave number $k$, for $l\gg1$ \cite{Oraevsky_WGM} 
\begin{equation}
	 kR \approx l + \beta_{\nu}\left( \frac{l}{2}\right)^{1/3} - P_{\sigma}, \label{WGM:Res} 
\end{equation} 
where $R$ is the radius of the sphere, $\beta_{\nu}\in\{2 . 3,4 . 1,5 . 5,\dots\}$ are the negative of the zeros of Airy's function $\mathrm{Ai}\left(x\right)$, $P_{\mathrm{TM}} = n_s /\sqrt{n_s^2 - 1}$, and $P_{\mathrm{TE}}^{- 1} = n_s\sqrt{n_s^2 - 1}$. $\mathcal{E}$ is a normalization constant chosen such that the integral over the system volume 
\begin{equation}
	 \int\left[ \frac{\epsilon_s}{2}|\mathbf{E}|^2 + \frac{1}{2\mu_0}|\mathbf{B}|^2\right] dV = \frac{\hbar\omega}{2}, 
\end{equation} 
where $i\omega\mathbf{B} = \nabla\times\mathbf{E}$, $\epsilon_s = \epsilon_0n_s^2$, and $\omega = kc/n_s$ with $n_s$ being the refractive index of the sphere. Then 
\begin{equation}
	 \mathcal{E} = \sqrt{\frac{\hbar\omega}{2\epsilon_sR^3\mathcal{N}_l(kR)}}, 
\end{equation} 
where 
\begin{align}
	 \mathcal{N}_l(x) & \overset{\triangle}{=}\int_0^1\tilde{r}^2 d\tilde{r}J_l^2\left( x\tilde{r}\right) \nonumber \\
	 & \approx \frac{J_l^2(x) - J_{l + 1}(x)J_{l - 1}(x)}{2}, \label{BessInt} 
\end{align} 
and the approximation holds again for $l\gg1$. The angular dependence for $l = m$ with $l\gg1$, \cite{Oraevsky_WGM} 
\begin{equation}
	 Y_l^l(\theta,\phi) \approx \left( \frac{l}{\pi}\right)^{1/4}\exp\left[ - \frac{l}{2}\left( \frac{\pi}{2} - \theta\right)^2\right] \frac{e^{il\phi}}{\sqrt{2\pi}}, \label{Yll:Gau} 
\end{equation} 
is a narrow Gaussian around $\theta = \pi/2$ with a width $\sqrt{2/l}$ and a traveling wave along the circle with wave number $l/R$. The radial dependence for $l\gg1$ \cite{AbrSteg} 
\begin{equation}
	 J_l(kr) \approx \left( \frac{2}{l}\right)^{1/3}\mathrm{Ai}\left( x - \beta_{\nu}\right) , \label{AiryApp} 
\end{equation} 
where the radial coordinate is scaled to 
\begin{equation}
	 x = \frac{l}{(l/2)^{1/3}}\left( 1 - \frac{r}{R}\right) . \label{RescalingBess} 
\end{equation} 

The leading interaction between magnons and WGMs is 2-photon 1-magnon scattering. Consider a TM polarized WGM $P\equiv\{p, - p^{\prime},\mu\}$ that scatters into a TE-polarized WGM $Q\equiv\{q,q^{\prime},\nu\}$ by absorbing a magnon $A$ (to be generalized below). We take in the following $p'>0$ and thus, back(forward) scattering corresponds to $q'>0$($q'<0$). The coupling constant depends on the modes as \cite{Wettling75,Borovik82}, 
\begin{equation}
	 G_{PQA} = \frac{n_s\epsilon_0\lambda_0}{\pi M_s}\int E_PE_Q^{\ast}\ \left( \Theta_CM_{A,\rho} - i\Theta_FM_{A,\phi}\right) dV, \label{Coup:Exp} 
\end{equation} 
where the integral is over the sphere's volume, $\lambda_0$ is the vacuum wavelength of the incident light, $M_s$ is the saturation magnetization, $\Theta_F$ is the Faraday rotation per unit length, $\Theta_C$ is the Cotton-Mouton ellipticity per unit length, and $M_{A,\phi}$($M_{A,\rho}$) is the $\phi$($\rho$)-component of $A$-magnons.  

For the uniform precession of the magnetization, i.e. the Kittel mode $K$, \cite{WalkerOrig}
\begin{equation}
	 M_{K,\phi} = iM_{K,\rho} = \sqrt{\frac{\hbar\gamma M_s}{2V_{\mathrm{sph}}}}, \label{Kitt} 
\end{equation} 
where $V_{\mathrm{sph}}$ is the volume of the sphere, and $\gamma$ is the modulus of the gyromagnetic ratio. We normalized the magnetization as 
\begin{equation}
	 \int\operatorname{Re}\left[ iM_{\phi}^{\ast}M_{\rho}\right] dV = \frac{\hbar\gamma M_s}{2}, \label{NormTwee} 
\end{equation} 
equivalent to Eq.~(\ref{Normalization}). The coupling constant is finite only when $q^{\prime} + p^{\prime} = 1$, $p - \left\vert p^{\prime}\right\vert = q - \left\vert q^{\prime}\right\vert $, and $\mu = \nu$ \cite{WGMOptoMag,JamesLowL}. The coupling constant, independent of optical modes,
\begin{equation}
	 \left\vert G_{PQK}\right\vert = G_K = \frac{c\left( \Theta_F + \Theta_C\right)}{n_s\sqrt{2sV_{\mathrm{sph}}}}, \label{CoupK} 
\end{equation} 
where $s = M_s/\gamma\hbar$ is the spin density. For the parameters in Table \ref{Tab:YIG}, $G_K = 2\pi\times 9.1\,\mathrm{Hz}$.  

An upper bound on $G_{PQA}$ for a given set of WGMs can be found by maximizing it over all normalized functions $\{M_{A,\rho}(\mathbf{r}),M_{A,\phi}(\mathbf{r})\}$. The solution $\mathbf{M}^{\mathrm{opt}}$ gives the magnetization profile with highest optomagnonic coupling. Later, we show that there exists eigenstates that are close to $\mathbf{M}^{\mathrm{opt}}$. We consider circularly polarized magnons $M_{A,\phi} = iM_{A,\rho}$ and discuss the effect of finite ellipticity below. By the method of Lagrange multipliers, 
\begin{equation}
	 \mathcal{L} = \int E_PE_Q^{\ast}M_{\phi}dV - \lambda\left( \int M_{\phi}^{\ast}M_{\phi}dV - \frac{\hbar\gamma M_s}{2}\right) 
\end{equation} 
is stationary at $M_{\phi} = M_{\phi}^{\mathrm{opt}}$. We find 
\begin{equation}
	 M_{\phi}^{\mathrm{opt}} = \frac{E_P^{\ast}E_Q}{\lambda}\propto J_p (k_Pr)J_q(k_Qr)Y_p^{p^{\prime}}Y_q^{q^{\prime}}, \label{OptMag} 
\end{equation} 
with 
\begin{equation}
	 \lambda = \sqrt{\frac{2}{\gamma\hbar M_s}\int\left\vert E_PE_Q\right\vert^2dV} . 
\end{equation} 
Therefore 
\begin{equation}
	 \mathcal{G}_{PQ}\overset{\triangle}{=}\left\vert G_{PQ,\mathrm{opt}}\right\vert = \frac{c\ \left( \Theta_F + \Theta_C\right)}{n_s \sqrt{2sV_{PQ}}}, \label{OptG} 
\end{equation} 
defining the effective overlap volume 
\begin{equation}
	 V_{PQ} = \frac{\left( \int\left\vert E_P\right\vert^2dV\right) \left( \int\left\vert E_Q\right\vert^2dV\right)}{\int\left\vert E_P \right\vert^2\left\vert E_Q\right\vert^2dV} . \label{VPQ} 
\end{equation} 
\begin{figure}[ptb]
\begin{equation*}
\includegraphics[width = \columnwidth,keepaspectratio]{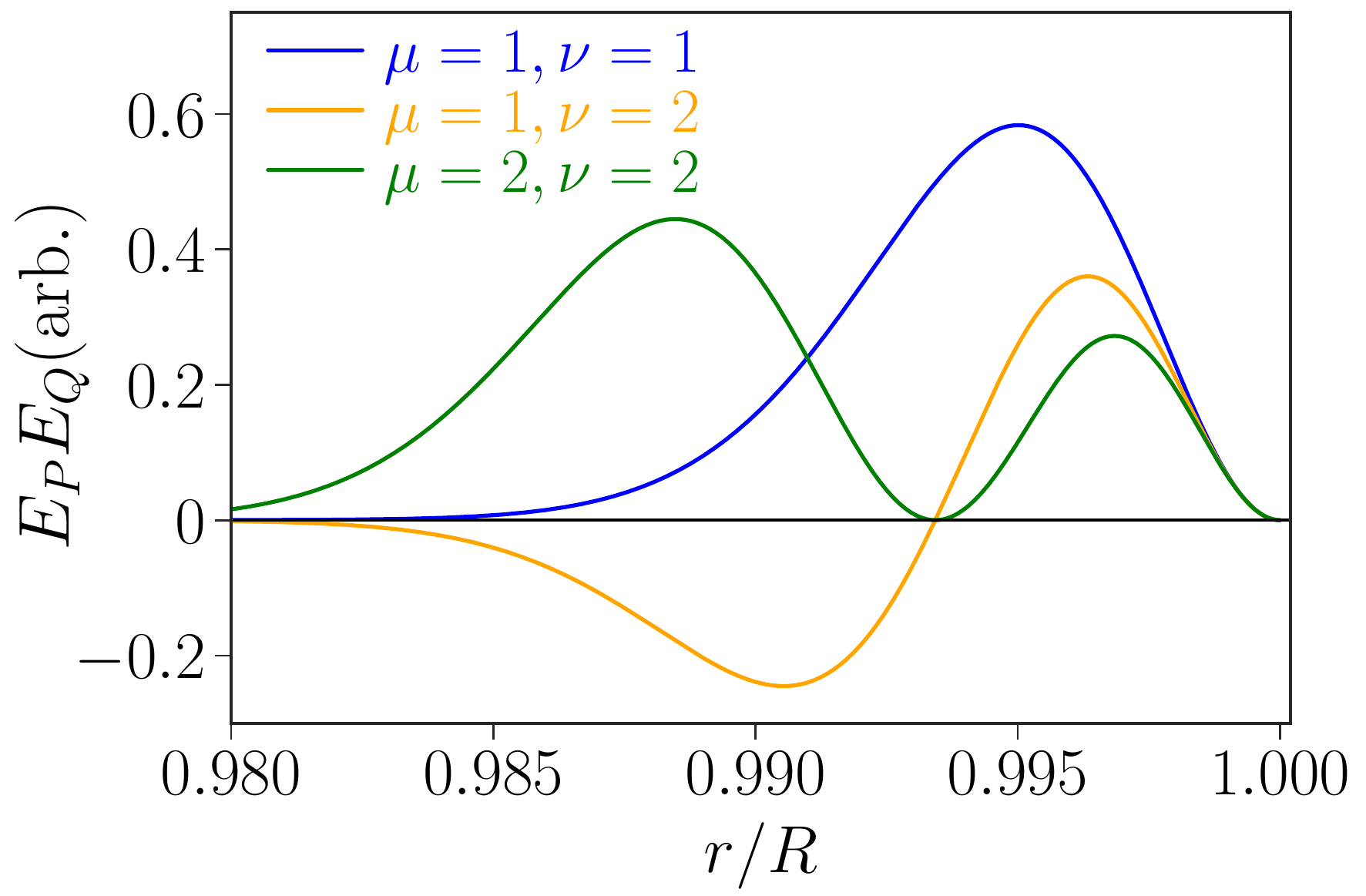}
\end{equation*}
\caption{The $r$-dependence of the product of the electric field of WGMs, in
arbitrary units, for $p=p^{\prime}=q=q^{\prime}=3000$ and radial mode numbers
$\mu,\nu\in\{1,2\}$. For the parameters of our system in Table \ref{Tab:YIG},
this corresponds to photons with free space wavelength $\approx
1.3\,\mathrm{\mu}$m. The magnons that match these profiles have the largest
optomagnonic coupling, cf. Eq.~(\ref{OptMag}). }
\label{Fig:Elec}
\end{figure}

The WGMs which are most concentrated to the surface have mode numbers $p = p^{\prime}$ and $q = q^{\prime}$. Since the magnon frequency $\sim1-10\,\text{GHz}$, is much smaller than that of the photons, $\sim200\,\text{THz}$, the incident and scattered photons have nearly the same frequency, implying $p \approx q$ [see Eq.~(\ref{WGM:Res})]. The Bessel function $J_p$ approaches the Airy function $\mathrm{Ai}(x)$ for $p,q\gg1$ [see Eq.~(\ref{AiryApp})], 
\begin{equation}
	 M_{\phi}^{\mathrm{opt}}\propto\mathrm{Ai}\left( x - \beta_{\mu}\right) \mathrm{Ai}\left( x - \beta_{\nu}\right) e^{- p\left( \frac{\pi}{2} - \theta\right)^2}e^{i(p + q)\phi}, \label{Mphi:PC} 
\end{equation} 
where the coordinate $x$ is given by Eq.~(\ref{RescalingBess}) after the substitution $l\rightarrow p$. This is a traveling wave in $\phi$-direction and a Gaussian in $\theta$-direction. Its radial dependence for the lowest $\{\mu,\nu\}$ is plotted in Fig.~\ref{Fig:Elec}, showing significant values only very close to the surface. The overlap volume (\ref{VPQ}) now reads 
\begin{equation}
	 V_{PQ} \approx \left( \frac{2}{p}\right)^{7/6}\frac{R^3\pi^{3/2}\left\vert \mathrm{Ai}^{\prime}\left( - \beta_{\mu}\right) \mathrm{Ai}^{\prime}\left( - \beta_{\nu}\right) \right\vert}{\int_0^{\infty}\mathrm{Ai}^2\left( x - \beta_{\mu}\right) \mathrm{Ai}^2\left( x - \beta_{\nu}\right) dx}, 
\end{equation} 
For $p = 3000$ and $\mu = \nu = 1$, $V_{\mathrm{sph}}/V_{PQ} \approx 1600$, reflecting the localized nature of the WGMs.  

For light with $\lambda_0 = 1 . 3\,\mathrm{\mu}$m, $p = 3190$ for a YIG sphere with parameters in Table \ref{Tab:YIG}. For the first modes $\left\{\mu ,\nu,\mathcal{G}_{PQ}/(2\pi)\right\} = \{1,1,364\,\text{Hz}\}$, $\{1,2,224\,\text{Hz}\}$, and $\{2,2,304\,\text{Hz}\},$ so $\mathcal{G}_{PQ}\gg G_K$. For a fixed $\lambda_0$, $p\propto R$, and $\mathcal{G}_{PQ}\propto R^{- 11/12}$ can be further enhanced by reducing the diameter.  

Magnetic anisotropies and dipolar interaction can deform the circular precession of the magnons into an ellipse. Solving the above problem for a hypothetical linearly polarized magnetization precession, e.g. by letting $M_{\phi}\rightarrow\infty$ and $M_{\rho}\rightarrow0$ while maintaining Eq.~(\ref{NormTwee}), leads to a diverging $\mathcal{G}_{PQ}\rightarrow\infty$. But such strong linear polarization are difficult to achieve in practice and ellipticity is typically limited to $\sim 10\%$, also valid in the calculations below.  

A similar analysis for $P$ and $Q$ being TE and TM polarized, respectively, reveals the same results with $\Theta_F + \Theta_C\rightarrow\Theta_F - \Theta_C$ and thus reduced couplings by a factor $0 . 45$. It is therefore advantageous to input TM photons over TE for larger blue sideband (magnon absorption) \cite{Sandercock73,Wettling75}. The coupling constant concerning magnon emission processes follows a very similar discussion since $G_{PQA}^{\mathrm{blue}} = G_{QPA}^{\ast}$.  

\begin{table}[ptb]
\begin{tabular}
[c]{p{.23\columnwidth}p{.23\columnwidth}p{.23\columnwidth}p{.23\columnwidth}}
\hline
$\lambda_{\mathrm{ex}}$ & $n_{s}$ & $M_{s}$ & $\gamma/(2\pi)$\\
$109 \text{nm}$ & $2.2$ & $140\ \text{kA/m}$ & $28\ \text{GHz/T}$\\\hline
$\Theta_{F}$ & $\Theta_{C} $ & $H_{\mathrm{app}}-M_{s}/3$ & $R$\\
$400\ \text{rad/m}$ & $150\ \text{rad/m}$ & $200\,\text{mT}/\mu_{0}$ &
$300$\thinspace$\mu$m\\\hline
\end{tabular}
\caption{Parameters for a standard YIG sphere: exchange constant
$A_{\mathrm{ex}}$ \cite{StanPrabh,KlinglerDex}, refractive index $n_{s}$
\cite{StanPrabh}, saturation magnetization $M_{s}$ \cite{StanPrabh},
gyromagnetic ratio $\gamma$ \cite{StanPrabh}, Faraday rotation angle
$\Theta_{F}$ \cite{Deeter90,ScottLacklison}, Cotton-Mouton ellipticity
$\Theta_{C}$ \cite{Castera77,Kamada_2001,James15}. We assume the applied dc
field $H_{\mathrm{app}}$ and the radius $R$ based on typical experimental
setup \cite{ZhangWGM16,Osada16,JamesWGM}. }
\label{Tab:YIG}
\end{table}

\section{Landau-Lifshitz equation} \label{Sec:LL}
  
Here we derive the equations for the magnetic eigenmodes which will later be shown to approximate the optimal profile derived above. The parameters for a standard YIG sphere are given in table \ref{Tab:YIG}. The Gilbert damping does not affect the magnon mode shapes to leading order and is disregarded. The magnetization dynamics then obeys the Landau-Lifshitz equation 
\begin{equation}
	 \frac{d\mathbf{M}}{dt} =  - \gamma\mu_0\mathbf{M}\times\mathbf{H}_{\mathrm{eff}}, \label{LL} 
\end{equation} 
where $\mathbf{M}$ is the magnetization, $\mu_0$ is the free space permeability, and the effective magnetic field 
\begin{equation}
	 \mathbf{H}_{\mathrm{eff}} = H_{\mathrm{app}}\hat{\mathbf{z}} + \frac{2A_{\mathrm{ex}}}{\mu_0M_s^2}\nabla^2\mathbf{M} + \mathbf{H}_{\mathrm{dip}}, 
\end{equation} 
where $H_{\mathrm{app}}$ is the applied field that saturates the magnetization to $M_s$ in the $\hat{\mathbf{z}}$-direction, $A_{ex}$ is the exchange constant, and $\mathbf{H}_{\mathrm{dip}}$ is the dipolar field that solves Maxwell's equations in the magnetostatic approximation: 
\begin{equation}
	 \nabla\times\mathbf{H}_{\mathrm{dip}} = 0;\ \ \nabla\cdot\mathbf{H}_{\mathrm{dip}} =  - \nabla\cdot\mathbf{M}, \label{Maxwell} 
\end{equation} 
which is valid for magnons with wavelengths sufficiently smaller than $c/\omega\sim1\,\text{cm}$ \cite{Hurben95}. The amplitudes $\mathbf{m} = \mathbf{M} - M_s\hat{\mathbf{z}}$ are taken to be small. The dipolar field has a large dc and a small ac component, $\mathbf{H}_{\mathrm{dip}} = \mathbf{H}_{\mathrm{demag}} + \mathbf{h}_{\mathrm{dip}},$ where the demagnetization field $\mathbf{H}_{\mathrm{demag}} =  - M_s\hat{\mathbf{z}}/3$ for a sphere. We disregard the small magneto-crystalline anisotropies in YIG.  

The scalar potential $\mathbf{h}_{\mathrm{dip}} =  - \nabla\psi$ satisfies 
\begin{equation}
	 \nabla^2\psi = \nabla\cdot\mathbf{m} . \label{PsiM} 
\end{equation} 
After substitution into Eq.~(\ref{LL}), linearizing in $\mathbf{m}$, and in the frequency domain $\partial/\partial t\rightarrow - i\omega$, 
\begin{equation}
	 \left[ \pm\omega + \omega_a - \frac{\omega_s}{k_{\mathrm{ex}}^2} \nabla^2\right] m_{\pm} =  - \omega_s\partial_{\pm}\psi, \label{LL:Lin} 
\end{equation} 
where we used the circular coordinates $m_{\pm} = m_x\pm im_y$ and $\partial_{\pm} = \partial_x\pm i\partial_y$. Here $\omega_a = \gamma \mu_0\left( H_{\mathrm{app}} - M_s/3\right) $, $\omega_s = \gamma \mu_0 M_s$, and the inverse exchange length 
\begin{equation}
	 \frac{2\pi}{\lambda_{\mathrm{ex}}} = k_{\mathrm{ex}} = \sqrt{\frac{\mu_0 M_s^2}{2A_{\mathrm{ex}}}} . 
\end{equation} 
We call $m_-$($m_+$) the Larmor(anti-Larmor) component since $m_+ = 0$ for a pure Larmor precession. Outside the magnet 
\begin{equation}
	 \nabla^2\psi_o = 0 . \label{Psiout} 
\end{equation} 
 
The coupled set of differential equations (\ref{PsiM})-(\ref{Psiout}) are closed by boundary conditions derived from Maxwell's equations at the interface,
\begin{equation}
	 \psi\left( R\right) = \psi_o\left( R\right) ;\ \ - \partial_r \psi(R) + m_r(R) =  - \partial_r\psi_o(R). \label{BD:Maxwell} 
\end{equation} 
The first condition is required for a finite $\mathbf{h}_{\mathrm{dip}}$ at the surface, while the second one enforces continuity of the normal component of the magnetic field $\mathbf{h}_{\mathrm{dip}} + \mathbf{m}$. At large distances, the magnetic field vanishes implying a constant potential which can be chosen to be zero,
\begin{equation}
	\psi_o\left(r \rightarrow \infty\right) = 0.
\end{equation}

The boundary conditions for the magnetization depends on the surface morphology and is complicated by the long range nature of the dipolar interaction \cite{SoohooBook,WamesWolfram_Detailed,Guslienko_BD}. Here, we present calculations for pinned boundary conditions, $m_{x,y}(R) = 0$, valid when the surface anisotropy is high \cite{SoohooBook,CamleyMills,Guslienko_BD} . This is not very realistic for samples with high surface quality but sufficiently accurate for our purposes, as justified in Sec. \ref{Sec:SolDisc}.  

\section{Exchange-dipole magnons} \label{Sec:SolDisc}
Here we discuss the amplitude of the magnons in dielectric magnetic spheres which resemble the ideal magnetization distribution derived in Sec. \ref{Sec:OMag}. These are the surface exchange-dipolar magnons localized at the equator derived in App. \ref{App:SolvingMagnons}. Similar problems have been addressed in Refs. \cite{WamesWolfram_Detailed,CylinderMagnons18} for different geometries.  

Analogous to the photons discussed above, magnons in spheres are characterized by three mode numbers $\{l,m,\nu\}$. Their amplitudes are a linear combination of three terms given in Eq.~(\ref{mag:Exp}) [cf. Eqs.~(\ref{mmin:App})-(\ref{mplu:App})] with `dispersion' relations in Eq.~(\ref{Dispersion}) [cf. Eq.~(\ref{DispApp})]. The partial waves appear with coefficients $\zeta$ defined below.
\begin{widetext}
\begin{equation}
	 m_{\pm}\left( \mathbf{r}\right) = m_0Y_{l\pm1}^{m\pm1}(\theta,\phi)\left[ \zeta_{\mathrm{dip},\pm}\left( \frac{r}{R}\right)^{l\pm1} + \zeta_{\mathrm{ex},\pm}\frac{J_{l\pm1}(kr)}{J_{l - 1}(kR)} + \zeta_{\mathrm{s},\pm}\frac{I_{l\pm1}(\kappa r)}{I_{l - 1}(\kappa R)}\right] . \label{mag:Exp}
\end{equation}
\begin{equation}
	 \frac{k^2}{k_{\mathrm{ex}}^2} = \frac{\omega_{\mathrm{sq}} - \omega_{\mathrm{DE}}}{\omega_s},\ \ \frac{\kappa^2}{k_{\mathrm{ex}}^2} = \frac{\omega_{\mathrm{sq}} + \omega_{\mathrm{DE}}}{\omega_s},\ \ \omega_{\mathrm{sq}} = \sqrt{\omega^2 + \frac{\omega_s^2}{4}},\ \ \omega_{\mathrm{DE}} = \omega_a + \frac{\omega_s}{2} . \label{Dispersion}
\end{equation}
\end{widetext}
Here $k_{\mathrm{ex}},\omega_s,\omega_a$ are defined below Eq.~(\ref{LL:Lin}), $\omega_{\mathrm{DE}}$ is the frequency of the surface magnons in a purely dipolar theory \cite{DamEshSlab,DamEshSurface}, and the normalization constant $m_0$ is determined below. \{`dip',`ex',`s'\} refers to \{dipolar, exchange, surface\} respectively. 

The ratios of anti-Larmor ($m_+$) and Larmor ($m_-$) components is a measure of the ellipticity [see Eq.~(\ref{LaLApp})]: 
\begin{equation}
	 \zeta_{\mathrm{dip} +} = 0,\ \ \frac{\zeta_{\mathrm{ex} +}}{\zeta_{\mathrm{ex} -}} = \frac{\omega_{\mathrm{sq}} - \omega}{\omega_s/2},\ \ \frac{\zeta_{\mathrm{s} +}}{\zeta_{\mathrm{s}, -}} = \frac{\omega_{\mathrm{sq}} + \omega}{\omega_s/2} . \label{LaLRatio} 
\end{equation} 
The coefficients $\zeta$ read for pinned boundary conditions $\mathbf{m}(R) = 0$ [see Eqs.~(\ref{zeta1})-(\ref{zeta2})], 
\begin{equation}
	 \zeta_{\mathrm{dip}, -} = \frac{\omega_{\mathrm{sq}}}{\omega_s/2} ,\ \ \zeta_{\mathrm{ex}, -} = \frac{- \kappa^2}{k_{\mathrm{ex}}^2} ,\ \ \zeta_{\mathrm{s}, -} = \frac{- k^2}{k_{\mathrm{ex}}^2} . \label{Coeff} 
\end{equation} 
Close to the boundary, the `dip' and `s' terms dominate, but the `ex' term in $m_{\pm}$ takes over for $r/R<1 - 1/l$.  

The dipolar (subscript `dip') term in Eq.~(\ref{mag:Exp}) decays exponentially with distance from the surface with a length scale $R/l$. This solution is not affected by exchange \cite{WalkerOrig,DamEshSurface} because $\nabla^2\left( Y_l^m(\theta,\phi)\left( \frac{r}{R}\right)^l\right) = 0$. For $l\gg1$ the surface term (subscript `s') simplifies by the asymptotics of the Bessel function to 
\begin{align}
	 \frac{I_{l - 1}(\kappa r)}{I_{l - 1}(\kappa R)} & \approx \left( \frac{\sqrt{l^2 + \kappa^2R^2} - l}{\sqrt{l^2 + \kappa^2R^2} + l}\right) \frac{I_{l + 1}(\kappa r)}{I_{l - 1}(\kappa R)} \nonumber \\
	 & \approx \exp\left[ - \sqrt{l^2 + \kappa^2R^2}\frac{R - r}{R}\right] . 
\end{align} 
This is again an exponential decay, but on an even shorter scale $R/\sqrt{l^2 + \kappa^2R^2}$ than the dipolar term. At first glance, it appears to have a large negative exchange energy, $\propto - \kappa^2$, but its total contribution to the energy is small due to its very small mode volume. Both `dip' and `s' terms are important to satisfy the boundary conditions, but they do not contribute significantly to the optomagnonic coupling because the optical WGMs penetrate much deeper into the magnet [see Fig. \ref{Fig:Elec}]. The exchange `ex' function in Eq.~(\ref{mag:Exp}), on the other hand, resembles a photon WGM when $kR \approx l$ [see Sec. \ref{Sec:OMag}]. We show below that this condition is satisfied by magnons with $\nu>0$.  

We now turn to the magnon eigenfrequencies and modes for fixed $l$ and $m$ with $\nu\geq0$ [using App. \ref{App:SolvingMagnons}]. For $\nu = 0$, $\omega_0^2 \approx \omega_a^2 + \omega_a\omega_s$ and mode amplitudes Eq.~(\ref{mag:Exp}) approach 
\begin{equation}
	 m_{\phi} \approx l^{3/2}\sqrt{\frac{\gamma\hbar M_s}{2R^3}}Y_l^m\left( \theta,\phi\right) \left( \frac{r}{R}\right)^{l - 1}\left( 1 - \frac{r^2}{R^2}\right) \label{Mag0} 
\end{equation} 
and $m_{\rho} =  - im_{\phi}$ when $k_{\mathrm{ex}}R\gg\sqrt{l}$, which is the case for typical experimental conditions discussed below. We normalized $m_{\phi}$ according to Eq.~(\ref{Normalization}). Note that (only) the results for $\nu = 0$ depend strongly on the surface pinning.  

For non-zero $\nu\sim O(1)$, analogous to Eq.~(\ref{WGM:Res}) for the photons, 
\begin{equation}
	 k_{\nu}R = l + \beta_{\nu}\left( \frac{l}{2}\right)^{1/3}, \label{MagRes} 
\end{equation} 
where $\beta_{\nu}\in\{2 . 3,4 . 1,5 . 5,\dots\}$ are again the negative of the zeros of Airy's function. We compute coefficients $\{\zeta_{\mathrm{dip} , -},\zeta_{\mathrm{ex}, -},\zeta_{\mathrm{s}, -},\zeta_{\mathrm{dip}, +} ,\zeta_{\mathrm{ex}, +}\} \approx \{3 . 5,3 . 4,0 . 1,0 . 5,1 . 0\}$. Although $\zeta_{\mathrm{ex}}\sim\zeta_{\mathrm{dip}}$, the energy of the `dip' term is much smaller than that of the `ex' term because the former is localized to a small skin depth $\sim R/l$ and therefore does not contribute much when integrated over the mode volume. We disregard `dip' and `s' terms at the cost of an error scaling as $\propto l^{- 1/3}$. The magnetization 
\begin{align}
	 m_{\phi}\left( \mathbf{r}\right) & \approx \sqrt{\frac{\gamma\hbar M_s}{2R^3\mathcal{N}_l(kR)}}Y_l^m(\theta,\phi)J_l(k_{\nu} r)\tan\theta_e\label{mag:mvp}\\
	 m_{\rho}\left( \mathbf{r}\right) & \approx  - im_{\phi}\left( \mathbf{r}\right) \cot^2\theta_e, \label{mag:mvp2}
\end{align} 
for $r/R<1 - 1/l$, where $\mathcal{N}$ is given by Eq.~(\ref{BessInt}). Since the magnetic field generated by magnetic dipoles is elliptically polarized, the magnetization precesses on an ellipse with major and minor axes along $\pmb{\rho}$ and $\pmb{\phi}$, respectively. The ellipticity is parametrized by the angle $\theta_e$, given by 
\begin{equation}
	 \tan\theta_e = \sqrt{\frac{\zeta_{\mathrm{ex}, -} - \zeta_{\mathrm{ex}, +}}{\zeta_{\mathrm{ex}, -} + \zeta_{\mathrm{ex}, +}}} = \sqrt{\frac{\omega_s /2 - \omega_{\mathrm{sq}} + \omega}{\omega_s/2 + \omega_{\mathrm{sq}} - \omega}} . \label{Ellip} 
\end{equation} 
The amplitudes (\ref{mag:mvp}) are normalized according to Eq.~(\ref{Normalization}). 

\begin{figure}[ptb]
\begin{equation*}
\includegraphics[width = \columnwidth,keepaspectratio]{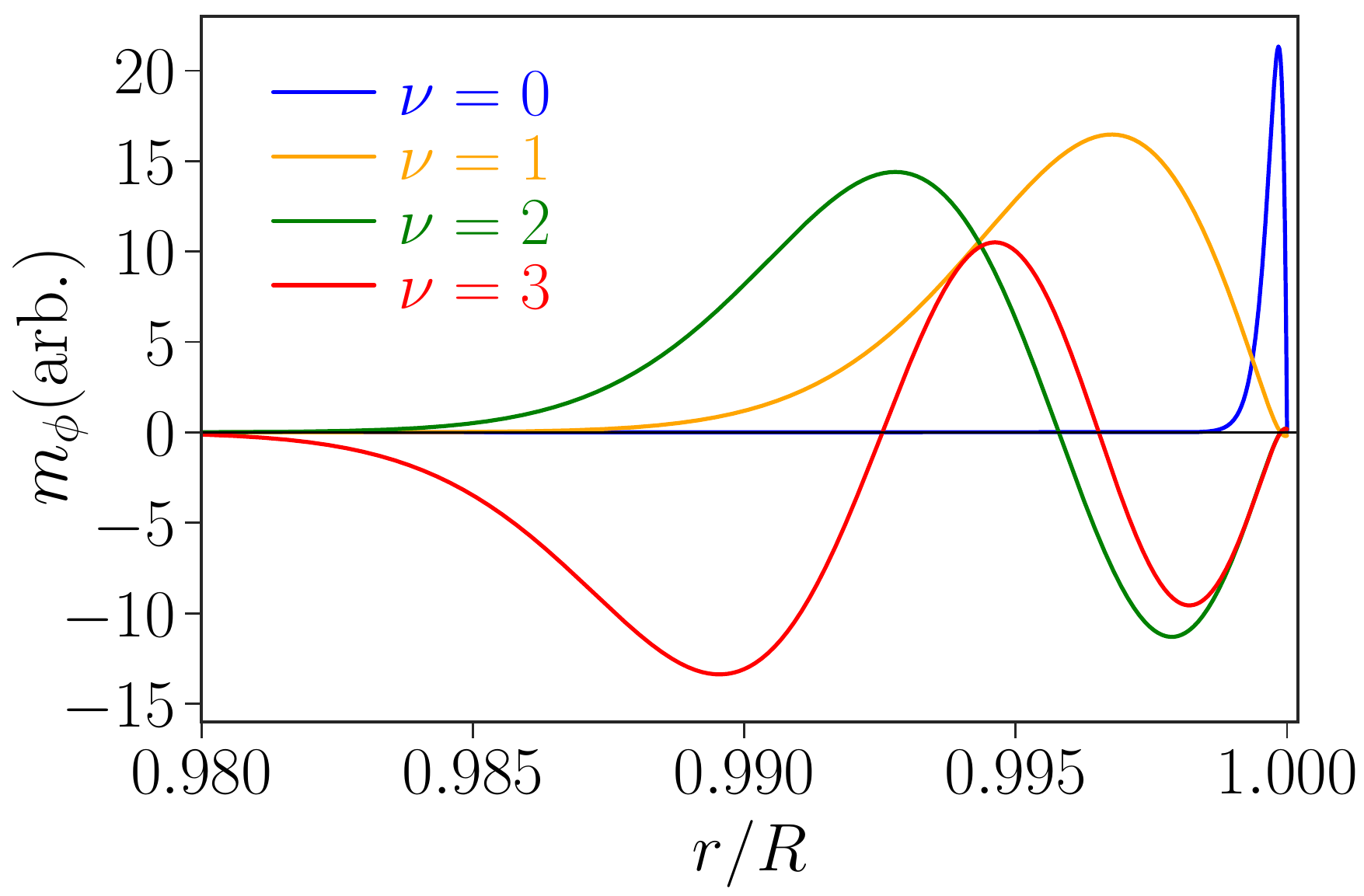}
\end{equation*}
\caption{Radial dependence of $m_{\phi}=(m_{+}e^{-i\phi}-m_{-}e^{i\phi})/2$
for $\nu\leq3$ and $l=6000$ with parameters from Table I. $\nu=0$ resembles a
purely dipolar wave and is localized to $1>r/R>1-2/l$. For $\nu>0$ the
magnetization is dominated by the Bessel function except for the region
occupied by the $\nu=0$ mode.}
\label{Fig:Mag}
\end{figure}

For $R = 300\,\mathrm{\mu}$m and $l = 6000$ [see Sec. \ref{Sec:Coup}], $2\pi R/l \approx 300\,$nm is the magnon wavelength for a typical experiment. The $\phi$-component of the magnetization $m_{\phi}$ for $\nu\leq3$ is plotted in Fig. \ref{Fig:Mag}, while $m_{\rho}$ looks similar to $m_{\phi}$ after scaling (not shown for brevity). $\nu>0$ modes contribute significantly to the coupling with large overlap factors [see Sec. \ref{Sec:Coup} for explicit expressions].  

For the parameters in Table \ref{Tab:YIG}, we find $\omega_a = 2\pi \times 5.6\,$GHz, and $\omega_s = 2\pi\times 4.9\,$GHz. Putting $kR = l$ in Eq.~(\ref{Dispersion}), we get the frequency $\omega_N = 2\pi\times 8.4\,$GHz. $\omega_0 = 2\pi\times 7.7\,$GHz, while frequencies for $\nu = \{1,2,3\}$ are $\omega_{\nu} = \omega_N + 2\pi\times\{7.5,13.2,17.9\}\,$MHz respectively. We estimate the linewidth of the magnons $\sim \alpha_G \omega_{\nu}$, in terms of the (geometry-independent) bulk Gilbert constant $\alpha_G = 10^{- 4}$ \cite{WuHoff,Chang14}. The frequency splittings are an order of magnitude larger than the typical line width, so the magnon resonances are well defined. The exchange mode has a small ellipticity $\tan\theta_e = 0 . 8 $.  

At these frequencies the `surface' term in Eq.~(\ref{mag:Exp}) has wavelengths $2\pi/\kappa_{\nu} \approx 60\,$nm. It decays much faster into the sphere than the wavelength of infrared light, $>500\,$nm in YIG, which validates our statements above.  

We assumed perfect pinning at the boundary, $m_{\pm}\left( R\right) = 0,$ which is realistic only when surface anisotropies are strong \cite{SoohooBook,WamesWolfram_Detailed,Guslienko_BD}. While Eqs.~(\ref{mag:Exp})-(\ref{LaLRatio}) do not depend on the boundary conditions, the relative weights of three waves, $\{\zeta_{\mathrm{dip, -}},\zeta_{\mathrm{ex, -}},\zeta_{\mathrm{s, -}}\}$ do. However, the validity of Eq.~(\ref{mag:mvp}) depends only on the fact that the energy is dominated by the Bessel function which still holds for imperfect pinning and $\nu>0$. We estimate the contributions of surface exchange waves to the magnon mode energy by the parameter 
\begin{equation}
	 \eta = \frac{\left\vert \zeta_{\mathrm{dip}, -}\right\vert^2}{\left\vert \zeta_{\mathrm{ex}, -}\right\vert^2}\frac{J_l^2(kR)\int(r/R)^{2l} dr}{\int J_l^2(kr)dr} . 
\end{equation} 
For a film, the squared ratio of the $\zeta$ coefficients is $\sim1$ \cite{WamesWolfram_Detailed}, which should be the case also for a sphere with curvature $R$ much larger than the magnon wavelength $R/l$. The second fraction is of $O(l^{- 1/3})$. Therefore $\eta\ll1,$ implying that the energy is indeed dominated by the Bessel function as assumed in Eq.~(\ref{mag:mvp}). Reduced pinning changes the magnetization profile near the surface, $r/R>1 - 1/l$, but not the coupling of states with $\nu>0$ to the WGMs.  

\section{Optomagnonic coupling} \label{Sec:Coup}

We calculate the coupling constant $G_{PQA}$ given by Eq.~(\ref{Coup:Exp}). Consider an incident TM-polarized optical WGM $P\equiv\{p, - p^{\prime},\mu\}$ that reflects into a TE-polarized WGM $Q\equiv\{q,q^{\prime},\nu\}$ by absorbing a magnon $A\equiv\{\alpha,\alpha^{\prime},\xi\}$. Their frequencies are, respectively, $\omega_P$, $\omega_Q$, and $\omega_A\ll\omega_P $,$\omega_Q$. By energy conservation, $\omega_P \approx \omega_Q$ and thus, $p \approx q$ [see Eq.~(\ref{WGM:Res})]. For the modes localized near the equator, $\theta = \pi/2$, the indices $x \approx x^{\prime}$ where $x\in\{p,q,\alpha\}$. The conservation of angular momentum in the $z$-direction \cite{WGMOptoMag}, cf. Eq.~(\ref{Oang:Res}), implies $p^{\prime} + q^{\prime} = \alpha^{\prime}$. For $\lambda_0 \approx 1 . 3\,\mathrm{\mu}$m, Eq.~(\ref{WGM:Res}) and Table \ref{Tab:YIG} give $p \approx 3000$ for $\nu_P\sim O(1)$. Summarizing, $p \approx p^{\prime} \approx q \approx q^{\prime} \approx \alpha/2 \approx \alpha^{\prime}/2 \approx 3000$.  

From Figs. \ref{Fig:Elec} and \ref{Fig:Mag}, we observe that the radial magnon amplitude can be close to the optimal profile. This is also the case in the azimuthal $\theta$-direction close to the equator (not shown). Here, we confirm this observation by explicitly calculating the mode overlap integrals.  

The coupling constant Eq.~(\ref{Coup:Exp}) can be written 
\begin{equation}
	 G_{PQA} = \frac{c(\Theta_F + \Theta_C)}{n_s\sqrt{2sR^3}}\mathcal{A}_{PQA}\mathcal{R}_{PQA}, \label{IntDivs} 
\end{equation} 
in terms of the dimensionless angular and radial overlap integrals, $\mathcal{A}_{PQA}$ and $\mathcal{R}_{PQA}$. 

The angular part, 
\begin{equation}
	 \mathcal{A}_{PQA} = \int Y_p^{- p^{\prime}}Y_{\alpha}^{\alpha^{\prime}}\left( Y_q^{q^{\prime}}\right)^{\ast}\sin\theta d\theta d\phi . 
\end{equation} 
is a standard integral that can be written in terms of Clebsch-Gordan coefficients $\left\langle l_1m_1,l_2m_2\middle|l_3m_3 \right\rangle $. For $p,q,\alpha\gg1$, 
\begin{equation}
	 \mathcal{A}_{PQA} \approx \sqrt{\frac{pq}{2\pi\alpha}}\left\langle pp^{\prime},qq^{\prime}\middle|\alpha\alpha^{\prime}\right\rangle \left\langle p0,q0\middle|\alpha0\right\rangle . 
\end{equation} 
 With $x = x^{\prime}$ where $x\in\{p,q,\alpha\}$, the Gaussian approximation [Eq.~(\ref{Yll:Gau})] leads to 
\begin{equation}
	 \mathcal{A}_{PQA} \approx \delta_{\alpha,p + q}\frac{(pq\alpha)^{1/4}}{\pi^{3/4}\sqrt{p + q + \alpha}} \approx \delta_{{\alpha},p + q}\frac{p^{1/4}}{3 . 97}, \label{Oang:Res} 
\end{equation} 
where in the second step, we used $p \approx q \approx{\alpha}/2$. $\mathcal{A}_{PQA}$ vanishes when ${\alpha}\neq p + q$, reflecting the conservation of angular momentum in the $z$-direction. The angular overlap is optimal because $Y_{\alpha}^{\alpha}\propto Y_p^pY_q^q$ for $p \approx q \approx \alpha/2,$ which equals the angular part in Eq.~(\ref{Mphi:PC}). For $p = 3000$, $\mathcal{A}_{PQA} = 1 . 9$.  

We discuss the radial overlap first for the magnon $\xi = 0$ with magnetization given by Eq.~(\ref{Mag0}). Then 
\begin{equation}
	 \mathcal{R}_{PQA}^{(0)} = \int_0^R\frac{\alpha^{3/2}J_p(k_Pr)J_q (k_Qr)}{\sqrt{\mathcal{N}_p(k_PR)\mathcal{N}_q(k_QR)}} \frac{r^{\alpha + 1}(R^2 - r^2)}{R^{\alpha + 4}}dr 
\end{equation} 
where $\{k_P,k_Q\}$ are the photon wave numbers, Eq.~(\ref{WGM:Res}). Since the magnetic amplitude is significant only near the surface, we may linearize the optical fields (the Bessel functions) close to $R$. Using Eq.~(\ref{WGM:Res}) and the Airy's function approximation \cite{AbrSteg} , cf. Eq.~(\ref{AiryApp}) 
\begin{equation}
	 J_p(k_Pr) \approx \frac{2^{2/3}\mathrm{Ai}^{\prime}\left( - \beta_{\mu}\right)}{p^{2/3}}\left[ P_{\mathrm{TM}} + p\left( 1 - \frac{r}{R}\right) \right] , 
\end{equation} 
and 
\begin{equation}
	 \mathcal{N}_p(k_PR) \approx \left( \frac{2}{p}\right)^{4/3} \frac{\mathrm{Ai}^{\prime2}\left( - \beta_{\mu}\right)}{2} . 
\end{equation} 
Similar results hold for $\{p,P,\mu,P_{\mathrm{TM}}\}\rightarrow \{q,Q,\nu,P_{\mathrm{TE}}\}$. For $p \approx q \approx{\alpha}/2$, 
\begin{equation}
	 \mathcal{R}_{PQA}^{(0)} = \sqrt{\frac{2}{p}}\left[ P_{\mathrm{TM}}P_{\mathrm{TE}} + P_{\mathrm{TM}} + P_{\mathrm{TE}} + \frac{3}{2}\right] . 
\end{equation} 
For $p = 3000$ and $n_s = 2 . 2$, $\mathcal{R}_{PQA}^{(0)} = 0 . 08$ and the coupling $G_{PQA}^{(0)} = 2\pi\times 2.8\,$Hz is of the same order as that to the Kittel mode, $G_K = 2\pi\times 9.1\,$Hz [see Sec. \ref{Sec:OMag}] \cite{WGMOptoMag}. We emphasize that this result depends strongly on the magnetic boundary condition (taken to be fully pinned here) and only indicates the smallness of the coupling.  

The magnetization Eq.~(\ref{mag:mvp}) for $\xi\geq1$ gives
\begin{equation}
	 \frac{\mathcal{R}_{PQA}}{M_e} \approx \int_0^R\frac{dr}{R}\frac{J_p(k_Pr)J_q(k_Qr)J_{\alpha}(k_Ar)}{\sqrt{\mathcal{N}_p (k_PR)\mathcal{N}_q(k_QR)\mathcal{N}_{\alpha}(k_AR)}}, \label{Or:Def} 
\end{equation} 
to leading order in $\alpha$, where 
\begin{equation}
	 M_e = \frac{\tan\theta_e\Theta_F + \cot\theta_e\Theta_C}{\Theta_F + \Theta_C} . \label{Ellipticity} 
\end{equation}
For a YIG sphere with parameters in table \ref{Tab:YIG}, the ellipticity of the magnons $\tan\theta_e = 0 . 8$ and $M_e \approx 0 . 95$. The parameter $M_e$ takes into account that $m_{\rho}$ and $m_{\phi}$ contribute differently to the coupling being proportional to the magneto-optical constants $\Theta_C$ and $\Theta_F$, respectively [see Eq.~(\ref{Coup:Exp})]. In YIG $\Theta_F > \Theta_C$ in the infrared [see Table~\ref{Tab:YIG}], so the coupling is reduced because $|m_{\phi}| < |m_{\rho}|$ [see Eqs.~(\ref{mag:mvp}) and (\ref{mag:mvp2})].

The Bessel functions asymptotically become Airy's functions, Eq.~(\ref{AiryApp}), 
\begin{equation}
	 \frac{\left\vert \mathcal{R}_{PQA}\right\vert}{M_e} \approx \sqrt{2} p^{1/3}\int_0^{\infty}\mathrm{A}_{\mu}\left( x\right) \mathrm{A}_{\nu}\left( x\right) \mathrm{A}_{\xi}\left( 2^{2/3}x\right) dx, \label{Orad:Res} 
\end{equation} 
where the scaled radial coordinate $x$ 
\begin{equation}
	 x = \frac{l}{(l/2)^{1/3}}\left( 1 - \frac{r}{R}\right),
\end{equation} 
and the normalized Airy's function, 
\begin{equation}
	 \mathrm{A}_o\left( x\right) = \frac{\mathrm{Ai}\left( x - \beta_o\right)}{\left\vert \mathrm{Ai}^{\prime}\left( - \beta_o\right) \right\vert} . 
\end{equation} 

$\mathcal{R}_{PQA}$ mainly depends on the radial structure of the mode amplitudes with a weak scaling factor of $p^{1/3}$. We summarize results as $\{\mu,\nu,\xi,\mathcal{R}_{PQA}\},$ where $\xi$ is chosen to maximize $\mathcal{R}_{PQA}$ for given $\{\mu,\nu\}$. For $p = 3000$, we find $\{1,1,1,8 . 02\}$, $\{1,2,1,3 . 64\}$, and $\{2,2,3,5 . 63\}$, much larger than the dipolar mode $\mathcal{R}^{(0)}_{PQA} = 0.08$.

\begin{table}[ptb]
\begin{tabular}
[c]{p{8mm}p{8mm}p{8mm}p{2cm}p{1cm}}\hline
$\mu$ & $\nu$ & $\xi$ & $\mathbb{G}_{PQ}/(2\pi)$ & $M_{r}$\\\hline
$1$ & $1$ & $1$ & $304$ & $0.88$\\\hline
$1$ & $2$ & $1$ & $138$ & $0.65$\\\hline
$2$ & $2$ & $3$ & $213$ & $0.74$\\\hline
$1$ & $3$ & $2$ & $144$ & $0.82$\\\hline
$2$ & $3$ & $4$ & $130$ & $0.66$\\\hline
$3$ & $3$ & $5$ & $180$ & $0.70$\\\hline
\end{tabular}
\caption{The calculated optomagnonic coupling for a given $\{\mu,\nu\}$ and $\xi$ chosen to maximize $G_{PQA}$. $M_{r}$ is the radial overlap defined in the text, such that $M_r = 1$ for the ideal magnetization distribution. $M_r \sim 1$ indicates high overlap.}
\label{Tab:Res}
\end{table}

For a given pair $(P,Q)$, we define $\mathbb{G}_{PQ}$ as the maximum over all $G_{PQA}$. With $x = x^{\prime}$ where $x\in\{p,q,\alpha\}$, the angular momentum of the magnon is fixed by the WGMs, see Eq.~(\ref{Oang:Res}). The radial index can be found by maximizing the integral appearing in Eq.~(\ref{Orad:Res}) by enumerating it for each $\xi$. The maximum appears at $\xi\sim O(1)$ for $\mu,\nu\sim O(1)$, so we do not need to go beyond $\xi = 10$.  

We present the final results in the table \ref{Tab:Res}, where $\mathbb{G}_{PQ}\sim2\pi\times200\,$Hz. This can be compared with the maximum coupling possible for WGMs, $\mathcal{G}_{PQ}$ discussed in Sec. \ref{Sec:OMag}. We find $\mathbb{G}_{PQ}/\mathcal{G}_{PQ} = M_eM_r$ where $M_e$ is given in Eq.~(\ref{Ellipticity}) and the radial `mismatch' 
\begin{equation}
	 M_r = \frac{2^{1/3}\int_0^{\infty}\mathrm{A}_{\mu}\left( x\right) \mathrm{A}_{\nu}\left( x\right) \mathrm{A}_{\xi}\left( 2^{2/3}x\right) dx}{\sqrt{\int_0^{\infty}\mathrm{A}_{\mu}^2\left( x\right) \mathrm{A}_{\nu}^2\left( x\right) dx}} . 
\end{equation} 
Table \ref{Tab:Res} indeed shows $M_r\sim O\left( 1\right) $ implying a near ideal mode matching. Furthermore, $\mathbb{G}_{PQ}\gg G_K$, the coupling to the Kittel mode. By doping with bismuth, the coupling can be increased tenfold \cite{LacklisonBiYIG} to $\mathbb{G}_{PQ}\sim2\pi \times\,2\,$kHz. We see that $\mathbb{G}_{PQ}/\mathcal{G}_{PQ}$ does not depend on $R$ and hence both scale $\mathbb{G}_{PQ},\mathcal{G}_{PQ}\propto R^{- 0 . 9}$. For a microsphere with $R = 10\,\mathrm{\mu}$m ($p \approx 100$), $\mathbb{G}_{PQ} \sim2\pi\times4\,$kHz is possible in YIG, but fabrication is challenging. A\ very similar theory as outlined here can be applied to YIG disks when their aspect ratio is close to unity and the demagnetization fields are approximately uniform. Scaling those down by nanofabrication of thin films may be the most straightforward option to enhance the coupling in otherwise monolithic optical wave guide structures.  

The above analysis for magnon cooling via $\text{TM}\rightarrow\text{TE}$ scattering can be generalized, similar to the discussion at the end of Sec. \ref{Sec:OMag}. The coupling constant $G_{\mathrm{TE}\rightarrow\mathrm{TM}}^{\mathrm{cool}}$ is smaller by a factor $\Theta_F - \Theta_C/(\Theta_F + \Theta_C) = 0 . 45$. Also, by Hermiticity, $\left\vert G_{\sigma\rightarrow\sigma^{\prime}}^{\mathrm{pump}}\right\vert = \left\vert G_{\sigma^{\prime}\rightarrow\sigma}^{\mathrm{cool}}\right\vert $ if the directions of motion are reversed as well.  

$A$-magnons are efficiently cooled by the process $P + A\rightarrow Q$ when the magnon annihilation rate exceeds that of the magnon equilibration. For the internal optical dissipation $\kappa_{\rm int}$ and the leakage rate of photons into the fiber $\kappa_{\rm ext}$, the cooperativity should satisfy \cite{OMagCool} 
\begin{equation}
	 C = \frac{4G_{PQA}^2n_P}{\left(\kappa_{\rm int} + \kappa_{\rm ext} \right)\kappa_A}>1 
\end{equation} 
where $n_P$ is the number of photons in $P$-mode, $\kappa_A\sim2\pi \times0 . 5$MHz is the magnon's linewidth in YIG, and $\kappa_{\rm int}\sim2\pi \times0 . 1 - 0 . 5\,$GHz \cite{ZhangWGM16,Osada16,JamesWGM}. We assumed $\omega_P + \omega_M = \omega_Q$ for simplicity. In terms of input power $P_{\rm in}$, \cite{OMagCool}
\begin{equation}
	n_P = \frac{4 \kappa_{\rm ext}}{\left(\kappa_{\rm int} + \kappa_{\rm ext} \right)^2} \frac{P_{\rm in}}{\hbar \omega_P}.
\end{equation}
The cooperativity $C$ is maximized at $\kappa_{\rm ext} = \kappa_{\rm int}/2$ for a given input power.

For $G_{PQA}\sim2\pi\times200\,$Hz, $C_{PQA} = 1$ for $n_P\sim10^9 - 10^{10}$ requiring large powers $P_{\rm in} \sim 50-1000$mW for $\omega_P = 2\pi \times 200$THz. However, required $P_{\rm in}$ can be significantly reduced by scaling or doping as discussed above: a tenfold increase in $G$ causes a hundredfold decrease in required input power. Similar arguments hold for magnon pumping processes $P\rightarrow A + Q^{\prime}$. The steady state number of magnons is governed by a balance of all cooling and pumping processes, whose analysis we defer to a future work.  

The strong coupling regime is reached under the condition $G_{PQA}\sqrt{n_P}> \left(\kappa_{\rm int} + \kappa_{\rm ext}\right),\kappa_A$ which again requires an unrealistically large $n_P>10^{12}$ for $G_{PQA}\sim2\pi\times200\,$Hz and powers exceeding kilowatts, because of the large optical linewidths observed in typical YIG spheres \cite{ZhangWGM16,Osada16,JamesWGM}. The optical lifetime is limited by material absorption \cite{ZhangWGM16} and thus, can be improved only at the cost of reduced magneto-optical coupling. 2-3 orders of magnitude improvement in coupling constant is required to bridge this gap.

\section{Discussion} \label{Sec:Disc}
  
We modeled the magnetization dynamics in spherical cavities in order to find its optimal coupling to WGM photons. We find that selected exchange-dipolar magnons localized close to the equator (but not the Damon-Eshbach modes) are almost ideally suited to play that role. We predict an up to $40$-fold increase in the coupling constant, implying a $1000$-fold larger signal in Brillouin light scattering, as compared to that of the (unexcited) Kittel mode. Further improvement requires smaller optical volumes or higher magneto-optical constants.  

The option to shrink the cavity and optical volume is limited by the wavelength $\lambda_0/n_s$. For $\lambda_0 = 1 . 3\,\mathrm{\mu}$m and $n_s = 2 . 2$, a cavity with an optical volume of $\lambda_0^3/n_s^3$ gives an upper limit $\sim2\pi\times50\,$kHz for pure YIG. In a Bi:YIG sphere of radius $\sim\lambda_0/n_s$, the optical first Mie resonance may strongly couple with the Kittel mode \cite{Evangelos}.

The coupling can be enhanced by the ellipticity angle $\theta_e$ of the magnetization, which is controlled by crystalline anisotropy, saturation magnetization, and geometry. Linear polarization $\theta_e\rightarrow0$ or $\theta_e\rightarrow\pi/2$ would lead to a diverging coupling, but in practice magnons are close to circularly polarized, $\theta_e \approx \pi/4$. For YIG spheres the weak ellipticity even suppresses the coupling, $M_e<1$ in Eq.~(\ref{Ellipticity}).  

In purely dipolar theory, the surface magnons are chiral, i.e. only modes with $m>0$ exist. Then, from Fig.~\ref{Fig:Sys}, magnon creation is not allowed leading to improved cooling of magnons \cite{OMagCool}. When the exchange interaction kicks in, propagation is not unidirectional \cite{Kostylev}, but we still expect suppression of the red sideband (magnon creation). We leave an analysis of the chirality of exchange-dipolar magnons to a future article.

We find that light may efficiently pump or cool certain surface (low wavelength) magnons that do not couple easily to microwaves. This could be used to manipulate macroscopically coherent magnons, raising hopes of accessing interesting non-classical dynamics in the foreseeable future.  

\begin{acknowledgments}
We thank T. Yu, S. Streib, M. Elyasi, and K. Sato for helpful input and discussions. This work is financially supported by the Nederlandse Organisatie voor Wetenschappelijk Onderzoek (NWO) as well as Grant-in-Aid for Scientific Research (Grant No. 26103006) of the Japan Society for the Promotion of Science (JSPS). 
\end{acknowledgments}

\appendix

\section{Exchange-dipolar magnons} \label{App:SolvingMagnons}
 Here, we solve Eqs.~(\ref{PsiM})-(\ref{Psiout}) with Maxwell boundary conditions, Eq.~(\ref{BD:Maxwell}), and pinned surface magnetization $m_{\pm}\left( R\right) = 0$. The magnetization in the linearized LL equation, Eq.~(\ref{LL:Lin}), can be eliminated in favor of the scalar potential $\psi$, Eq.~(\ref{PsiM}) \cite{WamesWolfram_Detailed}, 
\begin{equation}
	 \left[ \left( \mathcal{O}^2 - \omega^2\right) \nabla^2 + \omega_s\mathcal{O}\left( \nabla^2 - \frac{\partial^2}{\partial z^2}\right) \right] \psi = 0, \label{BessCubic} 
\end{equation} 
where $\mathcal{O} = \omega_a - D_{\mathrm{ex}}\nabla^2$ with $D_{\mathrm{ex}} = \omega_s/k_{\mathrm{ex}}^2$. The general solution for a sphere is complicated because the magnetization breaks the rotational symmetry, but it can be simplified for the surface magnons near the equator. The ansatz 
\begin{equation}
	 \psi(\mathbf{r}) = Y_l^m(\theta,\phi)\Psi(r), \label{psi2Psi}
\end{equation} 
where 
\begin{equation}
	 Y_l^m(\theta,\phi) = ( - 1)^m\sqrt{\frac{2l + 1}{4\pi}\frac{(l - m)!}{(l + m)!}}\ P_l^m(\cos\theta)e^{im\phi} \label{Def:Ylm} 
\end{equation} 
are spherical harmonic functions with associated Legendre polynomials 
\begin{equation}
	 P_l^m(x) = \frac{( - 1)^m}{2^ll!}\left( 1 - x^2\right)^{m/2} \frac{d^{l + m}}{dx^{l + m}}\left( x^2 - 1\right)^l, 
\end{equation} 
leads to $\nabla^2\psi = Y_l^m\hat{O}_l\Psi$ where 
\begin{equation}
	 \hat{O}_l = \frac{1}{r^2}\frac{\partial}{\partial r}\left( r^2 \frac{\partial}{\partial r}\right) - \frac{l(l + 1)}{r^2} 
\end{equation} 
have spherical Bessel functions of order $l$ as eigenfunctions. The surface magnons with large angular momentum $l$ are localized near the equator and have a large ``kinetic energy'' along the equator. The confinement along the $\theta$-direction is not so strong, however, so the magnon amplitude looks like a flat tire. A posteriori, we find $k_{\theta}\propto\sqrt{l},$ while $k_{\phi}\propto l$. For large $l$, the terms $\partial_z^2 \approx R^{- 2}\partial_{\theta}^2$ near the equator, may therefore be disregarded in Eq.~(\ref{BessCubic}). This gives a cubic in $\hat{O}_l$, similar to a magnetic cylinder \cite{CylinderMagnons18}, 
\begin{equation}
	 \hat{O}_l\left( \hat{O}_l + k^2\right) \left( \hat{O}_l - \kappa^2\right) \Psi = 0, \label{CubicEq} 
\end{equation} 
where 
\begin{equation}
	 D_{\mathrm{ex}}k^2 = \omega_{\mathrm{sq}} - \omega_a - \frac{\omega_s}{2},\ \ D_{\mathrm{ex}}\kappa^2 = \omega_{\mathrm{sq}} + \omega_a + \frac{\omega_s}{2}, \label{DispApp} 
\end{equation} 
where 
\begin{equation}
	 \omega_{\mathrm{sq}} = \sqrt{\omega^2 + \frac{\omega_s^2}{4}} . 
\end{equation} 
$\kappa$ is real and $k$ is real as well when $\omega>\sqrt{\omega_a^2 + \omega_a\omega_s}$, which is the case for $k \approx l/R$, i.e. waves propagating along the equator [see Sec. \ref{Sec:Coup}].  

Consider the eigenvalue equation $\hat{O}_l\Psi_{\mu} =  - \mu^2\Psi_{\mu}$ with reciprocal \textquotedblleft length scales\textquotedblright\ $\mu \in\{0,k,i\kappa\}$. Its two linearly independent solutions are spherical Bessel functions of first and second kind, which in the limit $l\gg1$ are proportional to Bessel functions of first [$J_l(\mu r)$] and second [$Y_l(\mu r),$ not to be confused with the spherical harmonic $Y_l^m$] kind, respectively. $Y_l(\mu r)$ diverges at $r = 0$, so inside the sphere $\Psi_{\mu} = J_l(\mu r)$. Thus, Eq.~(\ref{CubicEq}) has three linearly independent solutions, $\{\Psi_0,\Psi_k,\Psi_{i\kappa}\}$ and the general solution is
\begin{equation}
	 \Psi = \sum_{i = 1}^3\alpha_i\frac{J_l(\mu_ir)}{\mu_iJ_{l - 1}(\mu_iR)}, 
\end{equation} 
where $\mu_1\rightarrow0$, $\mu_2 = k$, $\mu_3 = i\kappa$, $\alpha_i$ are integration constants, and the Bessel functions 
\begin{equation}
	 J_l(z) = \sum_{r = 0}^{\infty}\frac{( - 1)^r}{r!(r + l)!}\left( \frac{z}{2}\right)^{2r + l} . \label{Def:Bess} 
\end{equation} 
The spatial distribution of the three components are discussed in more detail in the main text [see Sec. \ref{Sec:SolDisc}].  

Bringing back the angular dependence, $\psi = Y_l^m \Psi$ [see Eq.~(\ref{psi2Psi})], the derivative $\partial_{\pm} = \partial_x \pm i\partial_y$ (introduced in Sec. \ref{Sec:LL})
\begin{equation}
	 \partial_{\pm}\psi = Y_l^me^{\pm i\phi}\sum_{i = 1}^3\frac{\alpha_i}{J_{l - 1}(\mu_iR)}\left( J_l^{\prime}(\mu_ir)\mp\frac{mJ_l(\mu_i r)}{\mu_i\rho}\right) , 
\end{equation} 
where $\partial_{\pm} = \partial_x\pm i\partial_y$. Close to the equator, $\rho \approx r$ and using $l\gg|l - m|$, 
\begin{equation}
	 \partial_{\pm}\psi \approx \mp Y_{l\pm1}^{m\pm1}\sum_{i = 1}^3\frac{J_{l\pm 1}(\mu_ir)}{J_{l - 1}(\mu_iR)}, 
\end{equation} 
where we used the recursion relations \cite{AbrSteg} 
\begin{equation}
	 J_{\alpha\pm1}(x) = \frac{\alpha}{x}J_{\alpha}(x)\mp J_{\alpha}^{\prime}(x) 
\end{equation} 
and $Y_{l\pm1}^{m\pm1} \approx e^{\pm i\phi}Y_l^m$ that holds for $l\gg1,\left\vert l - m\right\vert $. Solving Eq.~(\ref{LL:Lin}) for magnetization, 
\begin{equation}
	 m_{\pm}\left( \mathbf{r}\right) = Y_{l\pm1}^{m\pm1}\sum_{i = 1}^3\zeta_{i,\pm}\frac{J_{l\pm1}(\mu_ir)}{J_{l - 1}(\mu_iR)}, \label{mpm:PreBD} 
\end{equation} 
with coefficients 
\begin{equation}
	 \zeta_{i,\pm} = \frac{\omega_s\alpha_i}{\omega\pm\tilde{\omega}_i}, \label{zetaApp} 
\end{equation} 
and $\tilde{\omega}_i = \omega_a + D_{\mathrm{ex}}\mu_i^2$.  

Outside the magnet, $\psi_o$ satisfies a Laplace equation Eq.~(\ref{Psiout}). Using the continuity of magnetic potential and $\psi_o\rightarrow0$ at $r\rightarrow\infty$, 
\begin{equation}
	 \psi_o = Y_l^m(\theta,\phi)\left( \frac{R}{r}\right)^{l + 1}\sum_{i = 1}^3\alpha_i\frac{J_l(\mu_iR)}{\mu_iJ_{l - 1}(\mu_iR)} . 
\end{equation}

The integration constants $\alpha_i$ are governed by the boundary conditions: Maxwell boundary conditions, Eq.~(\ref{BD:Maxwell}), and pinned magnetization boundary condition for the LL equation $m_{\pm} = 0,$ which\ we justified a posteriori in Sec. \ref{Sec:SolDisc}. Demanding $m_-(r = R) = 0$ and $\left . \partial_r(\psi - \psi_o)\right\vert_{r = R} = 0$ gives 
\begin{equation}
	 \sum_{i = 1}^3\frac{\omega_s\alpha_i}{\omega - \tilde{\omega}_i} = 0 = \sum_{i = 1}^3\alpha_i, \label{BDImpl} 
\end{equation} 
which is solved by 
\begin{align}
	 \alpha_1 &= m_0\frac{(\omega - \tilde{\omega}_1)(\tilde{\omega}_2 - \tilde{\omega}_3)}{\omega_s},\label{al1}\\
	 \alpha_2 &= m_0\frac{(\omega - \tilde{\omega}_2)(\tilde{\omega}_3 - \tilde{\omega}_1)}{\omega_s},\label{al2}\\ \alpha_3 &= m_0\frac{(\omega - \tilde{\omega}_3)(\tilde{\omega}_1 - \tilde{\omega}_2)}{\omega_s}, \label{al3} 
\end{align} 
where $m_0$ is a normalization constant.  

We now arrive at the solution discussed in the main text, Sec. \ref{Sec:SolDisc}. With $\{\mu_1,\mu_2,\mu_3\} = \{0,k,i\kappa\}$ 
\begin{equation}
	 \lim_{\mu_1\rightarrow0}J_l(\mu_1r) \approx \frac{1}{l!}\left( \frac{\mu_1r}{2}\right)^l;\ \ J_l(i\kappa r) = i^lI_l(\kappa r), 
\end{equation} 
where $I$ is the modified Bessel function. The above holds also for $l\rightarrow l\pm1$. Substituting into Eq.~(\ref{mpm:PreBD}), \begin{widetext}
\begin{alignat}{5}
	m_- &= Y_{l-1}^{m-1} \bigg[ \zeta_{1,-} \left(\frac{r}{R}\right)^{l-1} &&+ \zeta_{2,-} \frac{J_{l-1}(kr)}{J_{l-1}(kR)} &&+ \zeta_{3,-} \frac{I_{l-1}(\kappa r)}{I_{l-1}(\kappa R)} \bigg] \label{mmin:App} \\
	m_+ &= Y_{l+1}^{m+1} \bigg[ 0 &&+ \zeta_{2,+} \frac{J_{l+1}(kr)}{J_{l-1}(kR)} &&- \zeta_{3,+} \frac{I_{l+1}(\kappa r)}{I_{l-1}(\kappa R)} \bigg]. \label{mplu:App}
\end{alignat}
\end{widetext}

In spite of $J_{l - 1}(\mu_1r)\rightarrow0$, the first term of $m_-$ is finite while that of $m_+$ vanishes. The Bessel function ratios in the third terms are real even though $J_l(i\kappa r)$ need not be.  

According to Eq.~(\ref{zetaApp}) the polarization does not depend on the coefficients $\alpha_i$. With $\{\tilde{\omega}_1,\tilde{\omega}_2,\tilde{\omega}_3\} = \{\omega_a,\omega_{\mathrm{sq}} - \omega_s/2, - \omega_{\mathrm{sq}} - \omega_s/2\}$, $\omega_{\mathrm{sq}}^2 = \omega^2 + \omega_s^2/4$ 
\begin{equation}
	 \frac{\zeta_{2, +}}{\zeta_{2, -}} = \frac{\omega + \omega_s/2 - \omega_{\mathrm{sq}}}{\omega - \omega_s/2 + \omega_{\mathrm{sq}}} . \label{LaLApp} 
\end{equation} 
A similar result holds by substituting $\zeta_{2\pm}\rightarrow\zeta_{3\pm}$ and $\omega_{\mathrm{sq}}\rightarrow - \omega_{\mathrm{sq}}$. Multiplying the numerator and denominator in the above equation by $\omega - \omega_s /2 - \omega_{\mathrm{sq}}$, we arrive at the form Eq.~(\ref{LaLRatio}) in the main text.  

Substituting $\alpha_i$ for the pinned boundary conditions, Eqs.~(\ref{al1}-\ref{al3}), into Eq.~(\ref{zetaApp}) 
\begin{align}
	 \zeta_{1, -} &= m_0 \frac{2\omega_{\mathrm{sq}}}{\omega_s} \label{zeta1}\\
	 \zeta_{2, -} &=  - m_0\frac{\omega_a + \omega_{\mathrm{sq}} + \omega_s /2}{\omega_s},\label{zeta2}\\ \zeta_{3, -} &= m_0\frac{\omega_a - \omega_{\mathrm{sq}} + \omega_s /2}{\omega_s} . \label{zeta3} 
\end{align}

\begin{figure}[ptb]
\begin{equation*}
\includegraphics[width=\columnwidth,keepaspectratio]{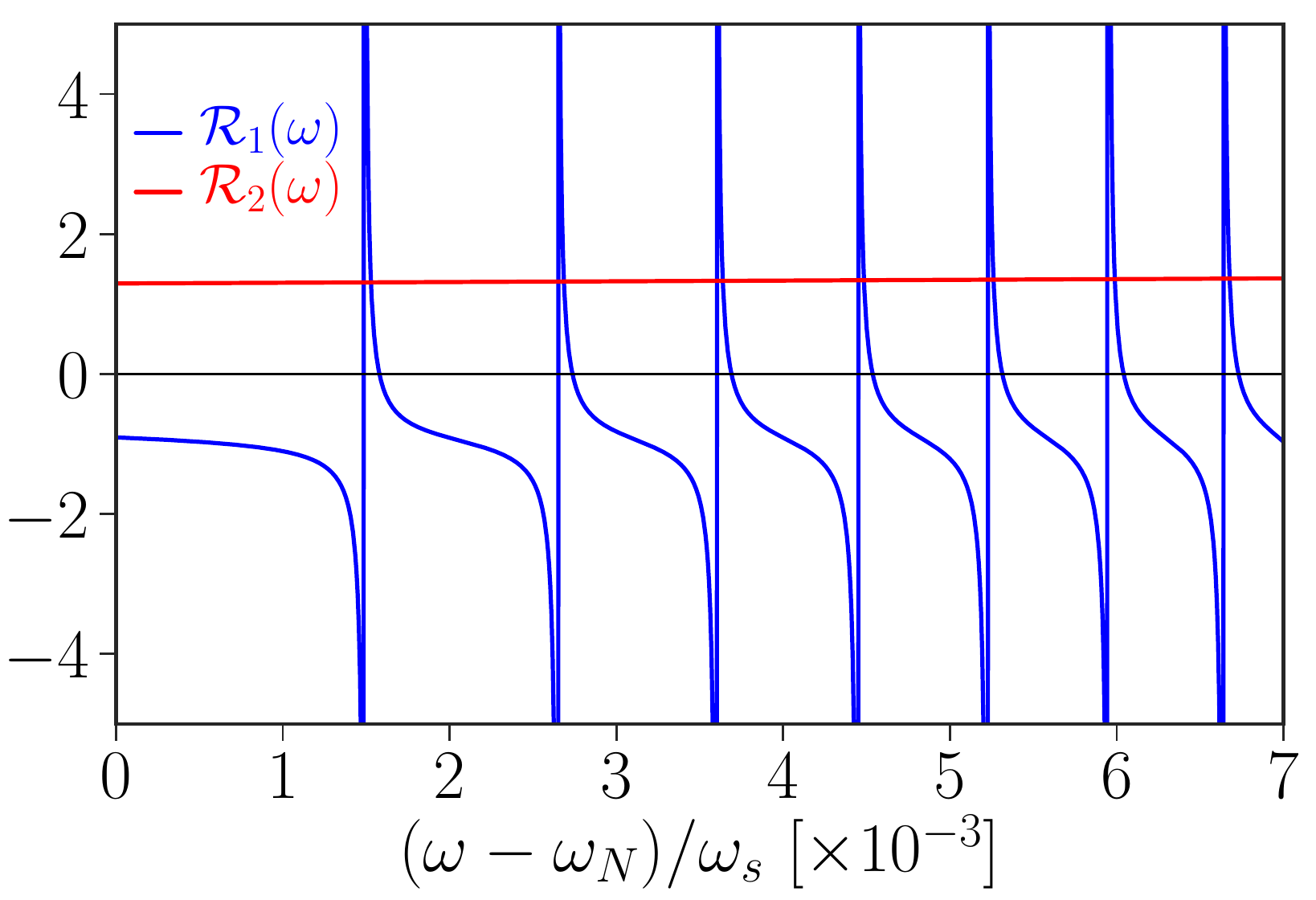}
\end{equation*}
\caption{The resonance condition $\mathcal{R}_{1}=\mathcal{R}_{2}$ gives the
allowed magnon frequencies when the magnetization is pinned at the surface.
$\omega_{N}$ is the frequency at which $kR=l$.}
\label{Fig:Res}
\end{figure}

\textit{\ }The above solutions satisfy Maxwell's boundary conditions, Eq.~(\ref{BD:Maxwell}), and $m_-(R) = 0$ by design [see Eq.~(\ref{BDImpl})]. The last condition $m_+(R) = 0$ gives the resonance condition $\mathcal{R}_1(\omega) = \mathcal{R}_2(\omega)$, where 
\begin{equation}
	 \mathcal{R}_1(\omega) =  - \frac{J_{l + 1}(kR)}{J_{l - 1}(kR)},\ \ \mathcal{R}_2(\omega) = \frac{k^2}{\kappa^2}\frac{\omega_{\mathrm{sq}} + \omega}{\omega_{\mathrm{sq}} - \omega}\frac{I_{l + 1}(\kappa R)}{I_{l - 1}(\kappa R)} . 
\end{equation} 
The roots of the above equation are counted by $\nu\geq0$. For $k>0$, the lowest root $\nu = 0$ occurs near $k \approx 0$ at frequency $\omega \approx \sqrt{\omega_a^2 + \omega_a\omega_s}$. The next and higher roots occurs only around $kR\gtrsim l$ as plotted in Fig. \ref{Fig:Res} [the root $\nu = 0$ is to the far left of the origin]. $\mathcal{R}_1$ is a rapidly varying function, while $\mathcal{R}_2 \approx 1 . 2$ is nearly constant. Sufficiently far from the zeroes of $J_{l - 1}(kR)$, $\mathcal{R}_1<0$ and at the crossing with $\mathcal{R}_2,$ $\mathcal{R}_1 \approx 1 . 2$. This implies that at magnon resonances, $J_{l - 1}(kR) \approx 0$ or $kR \approx l + \beta_{\nu}(l/2)^{1/3}$, while $\omega(k)$ is given by Eq.~(\ref{DispApp}). Their explicit values are discussed in Sec. \ref{Sec:SolDisc}  

\section{Normalization} \label{App:Norm}
 The classical Hamiltonian for a sphere that leads to the LL equation, Eq.~(\ref{LL}), reads \cite{StanPrabh} \textit{\ } 
\begin{equation}
	 H =  - \mu_0\int\left[ \left( H_{\mathrm{app}} - \frac{M_s}{3}\right) M_z + \frac{\mathbf{m}\cdot\mathbf{h}_{\mathrm{eff}}}{2}\right] dV, \label{MagHam} 
\end{equation} 
where 
\begin{equation}
	 \mathbf{h}_{\mathrm{eff}} = \frac{2A_{ex}}{\mu_0M_s^2}\nabla^2 \mathbf{m} + \mathbf{h}_{\mathrm{dip}} . 
\end{equation} 
and the integral is over all space. The solution of the linearized LL equation of motion gives a complete set of modes with spatiotemporal distribution $\mathbf{m}_p(\mathbf{r})e^{- i\omega_pt}$ and frequencies $\omega_p$. We may expand the fields 
\begin{equation}
	 A(\mathbf{r}) = \sum_{p,\omega_p>0}\left[ A_p(\mathbf{r})\alpha_p + A_p^{\ast}(\mathbf{r})\alpha_p^{\ast}\right] , 
\end{equation} 
where $A_p$ is the amplitude of any of $\{m_x,m_y,h_x,h_y\}$ of the $p$-th mode. Here and below the sum is restricted to positive frequencies. We have $\omega_a = \gamma\mu_0(H_{\mathrm{app}} - M_s/3)$, $\omega_s = \gamma\mu_0M_s$, and 
\begin{equation}
	 M_z \approx M_s - \frac{m_x^2 + m_y^2}{2M_s} . 
\end{equation} 
Eq.~(\ref{LL}) relates $\mathbf{m}_p$ and $\mathbf{h}_p$, 
\begin{align}
	 \omega_sh_{x,p} &= \omega_am_{x,p} + i\omega_pm_{y,p}\label{hm1}\\
	 \omega_sh_{y,p} &= \omega_am_{y,p} - i\omega_pm_{x,p} \label{hm2} 
\end{align} 
Inserting these into the Hamiltonian , 
\begin{equation}
	 H = \frac{\mu_0}{2}\sum_{pq}\left[ X_{pq}\alpha_p\alpha_q + X_{pq}^{\ast}\alpha_p^{\ast}\alpha_q^{\ast} + Y_{pq}\alpha_p\alpha_q^{\ast} + Y_{pq}^{\ast}\alpha_p^{\ast}\alpha_q\right] , 
\end{equation} 
where 
\begin{align}
	 X_{pq} &= \frac{i\omega_q}{\omega_s}\int\left( m_{y,p}m_{x,q} - m_{x,p}m_{y,q}\right) dV\\
	 Y_{pq} &= \frac{i\omega_q}{\omega_s}\int\left( m_{x,p}m_{y,q}^{\ast} - m_{y,p}m_{x,q}^{\ast}\right) dV . 
\end{align}

Following Ref. \cite{WalkerOrig}, we find orthogonality relations between magnons. For $\mathbf{b}_p = \mathbf{h}_p + \mathbf{m}_p,$ $\boldsymbol{\nabla}\cdot\mathbf{b}_p = 0$ from Maxwell's equations and 
\begin{equation}
	 \int\psi_q^{\ast}\boldsymbol{\nabla}\cdot\mathbf{b}_pdV = 0, \label{FirstOrth} 
\end{equation} 
where the scalar potential $\psi_q$ obeys $\nabla^2\psi_q = \nabla \cdot\mathbf{m}_q$. Integrating by parts and using $\mathbf{h}_q^{\ast} =  - \nabla\psi_q^{\ast}$, 
\begin{equation}
	 \int\left( \mathbf{h}_p + \mathbf{m}_p\right) \cdot\mathbf{h}_q^{\ast}dV = 0 . 
\end{equation} 
Using the same relation with $p\leftrightarrow q$ and subtracting, 
\begin{equation}
	 \int\left( \mathbf{m}_p\cdot\mathbf{h}_q^{\ast} - \mathbf{m}_q^{\ast}\cdot\mathbf{h}_p\right) dV = 0 . 
\end{equation} 
Substituting the mode-dependent fields $\mathbf{h}_{p(q)}$ from Eqs.~(\ref{hm1})-(\ref{hm2}), we find that $(\omega_p - \omega_q)Y_{pq} = 0$. A similar calculation starting with $\psi_q^{\ast}\rightarrow\psi_q$ in Eq.~(\ref{FirstOrth}) gives $(\omega_p + \omega_q)X_{pq} = 0$. Exchange breaks the degeneracy of the surface modes, as discussed in App. \ref{App:SolvingMagnons} . Since $\omega_p>0$, we conclude that $X_{pq} = 0$ and $Y_{pq}\propto\delta_{pq}$. The Hamiltonian is then reduced to that of a collection of harmonic oscillators: 
\begin{equation}
	 H = \mu_0\sum_pY_{pp}\left\vert \alpha_p\right\vert^2, 
\end{equation} 
where we used $Y_{pp} = Y_{pp}^{\ast}$.  

$\alpha_p$ is proportional to the amplitude of a magnon mode $p$. Correspondence with the quantum Hamiltonian for harmonic oscillators is achieved with a normalization that associates $\left\vert \alpha_p\right\vert^2$ to the number of magnons by demanding $\mu_0 Y_{pp} = \hbar\omega_p$ or 
\begin{equation}
	 \int\left( |m_{- ,p}|^2 - |m_{+ ,p}|^2\right) dV = 2\hbar\gamma M_s . \label{Normalization} 
\end{equation} 
For a pure (circular) Larmor precession, i.e. $m_+ = 0$, this condition can also be derived by assuming that the magnon has a spin of $\hbar$ since 
\begin{equation}
	 S_z = \int dV\frac{M_s - M_z}{\gamma} = \hbar\sum_p|\alpha_p |^2 . 
\end{equation} 
Vice versa, the spin of a magnon is not $\hbar$ when the precession is elliptic ($m_+\neq0$) \cite{AkashSqueezing}.  


\begin{thebibliography}{63}%
\makeatletter
\providecommand \@ifxundefined [1]{%
 \@ifx{#1\undefined}
}%
\providecommand \@ifnum [1]{%
 \ifnum #1\expandafter \@firstoftwo
 \else \expandafter \@secondoftwo
 \fi
}%
\providecommand \@ifx [1]{%
 \ifx #1\expandafter \@firstoftwo
 \else \expandafter \@secondoftwo
 \fi
}%
\providecommand \natexlab [1]{#1}%
\providecommand \enquote  [1]{``#1''}%
\providecommand \bibnamefont  [1]{#1}%
\providecommand \bibfnamefont [1]{#1}%
\providecommand \citenamefont [1]{#1}%
\providecommand \href@noop [0]{\@secondoftwo}%
\providecommand \href [0]{\begingroup \@sanitize@url \@href}%
\providecommand \@href[1]{\@@startlink{#1}\@@href}%
\providecommand \@@href[1]{\endgroup#1\@@endlink}%
\providecommand \@sanitize@url [0]{\catcode `\\12\catcode `\$12\catcode
  `\&12\catcode `\#12\catcode `\^12\catcode `\_12\catcode `\%12\relax}%
\providecommand \@@startlink[1]{}%
\providecommand \@@endlink[0]{}%
\providecommand \url  [0]{\begingroup\@sanitize@url \@url }%
\providecommand \@url [1]{\endgroup\@href {#1}{\urlprefix }}%
\providecommand \urlprefix  [0]{URL }%
\providecommand \Eprint [0]{\href }%
\providecommand \doibase [0]{http://dx.doi.org/}%
\providecommand \selectlanguage [0]{\@gobble}%
\providecommand \bibinfo  [0]{\@secondoftwo}%
\providecommand \bibfield  [0]{\@secondoftwo}%
\providecommand \translation [1]{[#1]}%
\providecommand \BibitemOpen [0]{}%
\providecommand \bibitemStop [0]{}%
\providecommand \bibitemNoStop [0]{.\EOS\space}%
\providecommand \EOS [0]{\spacefactor3000\relax}%
\providecommand \BibitemShut  [1]{\csname bibitem#1\endcsname}%
\let\auto@bib@innerbib\@empty
\bibitem [{\citenamefont {Chumak}\ \emph {et~al.}(2015)\citenamefont {Chumak},
  \citenamefont {Vasyuchka}, \citenamefont {Serga},\ and\ \citenamefont
  {Hillebrands}}]{Chumak15}%
  \BibitemOpen
  \bibfield  {author} {\bibinfo {author} {\bibfnamefont {A.~V.}\ \bibnamefont
  {Chumak}}, \bibinfo {author} {\bibfnamefont {V.~I.}\ \bibnamefont
  {Vasyuchka}}, \bibinfo {author} {\bibfnamefont {A.~A.}\ \bibnamefont
  {Serga}}, \ and\ \bibinfo {author} {\bibfnamefont {B.}~\bibnamefont
  {Hillebrands}},\ }\href {http://dx.doi.org/10.1038/nphys3347} {\bibfield
  {journal} {\bibinfo  {journal} {Nat Phys}\ }\textbf {\bibinfo {volume}
  {11}},\ \bibinfo {pages} {453} (\bibinfo {year} {2015})}\BibitemShut
  {NoStop}%
\bibitem [{\citenamefont {Cornelissen}\ \emph {et~al.}(2015)\citenamefont
  {Cornelissen}, \citenamefont {Liu}, \citenamefont {Duine}, \citenamefont
  {Youssef},\ and\ \citenamefont {van Wees}}]{Cornelissen15}%
  \BibitemOpen
  \bibfield  {author} {\bibinfo {author} {\bibfnamefont {L.~J.}\ \bibnamefont
  {Cornelissen}}, \bibinfo {author} {\bibfnamefont {J.}~\bibnamefont {Liu}},
  \bibinfo {author} {\bibfnamefont {R.~A.}\ \bibnamefont {Duine}}, \bibinfo
  {author} {\bibfnamefont {J.~B.}\ \bibnamefont {Youssef}}, \ and\ \bibinfo
  {author} {\bibfnamefont {B.~J.}\ \bibnamefont {van Wees}},\ }\href {\doibase
  10.1038/nphys3465} {\bibfield  {journal} {\bibinfo  {journal} {Nature
  Physics}\ }\textbf {\bibinfo {volume} {11}},\ \bibinfo {pages} {1022}
  (\bibinfo {year} {2015})}\BibitemShut {NoStop}%
\bibitem [{\citenamefont {Tabuchi}\ \emph {et~al.}(2016)\citenamefont
  {Tabuchi}, \citenamefont {Ishino}, \citenamefont {Noguchi}, \citenamefont
  {Ishikawa}, \citenamefont {Yamazaki}, \citenamefont {Usami},\ and\
  \citenamefont {Nakamura}}]{Tabuchi16}%
  \BibitemOpen
  \bibfield  {author} {\bibinfo {author} {\bibfnamefont {Y.}~\bibnamefont
  {Tabuchi}}, \bibinfo {author} {\bibfnamefont {S.}~\bibnamefont {Ishino}},
  \bibinfo {author} {\bibfnamefont {A.}~\bibnamefont {Noguchi}}, \bibinfo
  {author} {\bibfnamefont {T.}~\bibnamefont {Ishikawa}}, \bibinfo {author}
  {\bibfnamefont {R.}~\bibnamefont {Yamazaki}}, \bibinfo {author}
  {\bibfnamefont {K.}~\bibnamefont {Usami}}, \ and\ \bibinfo {author}
  {\bibfnamefont {Y.}~\bibnamefont {Nakamura}},\ }\href {\doibase
  http://dx.doi.org/10.1016/j.crhy.2016.07.009} {\bibfield  {journal} {\bibinfo
   {journal} {Comptes Rendus Physique}\ }\textbf {\bibinfo {volume} {17}},\
  \bibinfo {pages} {729 } (\bibinfo {year} {2016})}\BibitemShut {NoStop}%
\bibitem [{\citenamefont {Cherepanov}\ \emph {et~al.}(1993)\citenamefont
  {Cherepanov}, \citenamefont {Kolokolov},\ and\ \citenamefont
  {L'vov}}]{Cherepanov_YIG}%
  \BibitemOpen
  \bibfield  {author} {\bibinfo {author} {\bibfnamefont {V.}~\bibnamefont
  {Cherepanov}}, \bibinfo {author} {\bibfnamefont {I.}~\bibnamefont
  {Kolokolov}}, \ and\ \bibinfo {author} {\bibfnamefont {V.}~\bibnamefont
  {L'vov}},\ }\href {\doibase http://dx.doi.org/10.1016/0370-1573(93)90107-O}
  {\bibfield  {journal} {\bibinfo  {journal} {Physics Reports}\ }\textbf
  {\bibinfo {volume} {229}},\ \bibinfo {pages} {81 } (\bibinfo {year}
  {1993})}\BibitemShut {NoStop}%
\bibitem [{\citenamefont {Wu}\ and\ \citenamefont {Hoffmann}(2013)}]{WuHoff}%
  \BibitemOpen
  \bibfield  {author} {\bibinfo {author} {\bibfnamefont {M.}~\bibnamefont
  {Wu}}\ and\ \bibinfo {author} {\bibfnamefont {A.}~\bibnamefont {Hoffmann}},\
  }\href@noop {} {\emph {\bibinfo {title} {Recent Advances in Magnetic
  Insulators - From Spintronics to Microwave Applications}}},\ Vol.~\bibinfo
  {volume} {64}\ (\bibinfo  {publisher} {Elsevier},\ \bibinfo {year} {2013})\
  pp.\ \bibinfo {pages} {1--408}\BibitemShut {NoStop}%
\bibitem [{\citenamefont {Soykal}\ and\ \citenamefont
  {Flatte}(2010)}]{SoykalPRL10}%
  \BibitemOpen
  \bibfield  {author} {\bibinfo {author} {\bibfnamefont {O.~O.}\ \bibnamefont
  {Soykal}}\ and\ \bibinfo {author} {\bibfnamefont {M.~E.}\ \bibnamefont
  {Flatte}},\ }\href {\doibase 10.1103/PhysRevLett.104.077202} {\bibfield
  {journal} {\bibinfo  {journal} {Phys. Rev. Lett.}\ }\textbf {\bibinfo
  {volume} {104}},\ \bibinfo {pages} {077202} (\bibinfo {year}
  {2010})}\BibitemShut {NoStop}%
\bibitem [{\citenamefont {Huebl}\ \emph {et~al.}(2013)\citenamefont {Huebl},
  \citenamefont {Zollitsch}, \citenamefont {Lotze}, \citenamefont {Hocke},
  \citenamefont {Greifenstein}, \citenamefont {Marx}, \citenamefont {Gross},\
  and\ \citenamefont {Goennenwein}}]{HueblPol}%
  \BibitemOpen
  \bibfield  {author} {\bibinfo {author} {\bibfnamefont {H.}~\bibnamefont
  {Huebl}}, \bibinfo {author} {\bibfnamefont {C.~W.}\ \bibnamefont
  {Zollitsch}}, \bibinfo {author} {\bibfnamefont {J.}~\bibnamefont {Lotze}},
  \bibinfo {author} {\bibfnamefont {F.}~\bibnamefont {Hocke}}, \bibinfo
  {author} {\bibfnamefont {M.}~\bibnamefont {Greifenstein}}, \bibinfo {author}
  {\bibfnamefont {A.}~\bibnamefont {Marx}}, \bibinfo {author} {\bibfnamefont
  {R.}~\bibnamefont {Gross}}, \ and\ \bibinfo {author} {\bibfnamefont
  {S.~T.~B.}\ \bibnamefont {Goennenwein}},\ }\href {\doibase
  10.1103/PhysRevLett.111.127003} {\bibfield  {journal} {\bibinfo  {journal}
  {Phys. Rev. Lett.}\ }\textbf {\bibinfo {volume} {111}},\ \bibinfo {pages}
  {127003} (\bibinfo {year} {2013})}\BibitemShut {NoStop}%
\bibitem [{\citenamefont {Tabuchi}\ \emph {et~al.}(2014)\citenamefont
  {Tabuchi}, \citenamefont {Ishino}, \citenamefont {Ishikawa}, \citenamefont
  {Yamazaki}, \citenamefont {Usami},\ and\ \citenamefont
  {Nakamura}}]{TabuchiHybrid14}%
  \BibitemOpen
  \bibfield  {author} {\bibinfo {author} {\bibfnamefont {Y.}~\bibnamefont
  {Tabuchi}}, \bibinfo {author} {\bibfnamefont {S.}~\bibnamefont {Ishino}},
  \bibinfo {author} {\bibfnamefont {T.}~\bibnamefont {Ishikawa}}, \bibinfo
  {author} {\bibfnamefont {R.}~\bibnamefont {Yamazaki}}, \bibinfo {author}
  {\bibfnamefont {K.}~\bibnamefont {Usami}}, \ and\ \bibinfo {author}
  {\bibfnamefont {Y.}~\bibnamefont {Nakamura}},\ }\href {\doibase
  10.1103/PhysRevLett.113.083603} {\bibfield  {journal} {\bibinfo  {journal}
  {Phys. Rev. Lett.}\ }\textbf {\bibinfo {volume} {113}},\ \bibinfo {pages}
  {083603} (\bibinfo {year} {2014})}\BibitemShut {NoStop}%
\bibitem [{\citenamefont {Zhang}\ \emph {et~al.}(2014)\citenamefont {Zhang},
  \citenamefont {Zou}, \citenamefont {Jiang},\ and\ \citenamefont
  {Tang}}]{Zhang14}%
  \BibitemOpen
  \bibfield  {author} {\bibinfo {author} {\bibfnamefont {X.}~\bibnamefont
  {Zhang}}, \bibinfo {author} {\bibfnamefont {C.-L.}\ \bibnamefont {Zou}},
  \bibinfo {author} {\bibfnamefont {L.}~\bibnamefont {Jiang}}, \ and\ \bibinfo
  {author} {\bibfnamefont {H.~X.}\ \bibnamefont {Tang}},\ }\href {\doibase
  10.1103/PhysRevLett.113.156401} {\bibfield  {journal} {\bibinfo  {journal}
  {Phys. Rev. Lett.}\ }\textbf {\bibinfo {volume} {113}},\ \bibinfo {pages}
  {156401} (\bibinfo {year} {2014})}\BibitemShut {NoStop}%
\bibitem [{\citenamefont {Cao}\ \emph {et~al.}(2015)\citenamefont {Cao},
  \citenamefont {Yan}, \citenamefont {Huebl}, \citenamefont {Goennenwein},\
  and\ \citenamefont {Bauer}}]{YunshanPol}%
  \BibitemOpen
  \bibfield  {author} {\bibinfo {author} {\bibfnamefont {Y.}~\bibnamefont
  {Cao}}, \bibinfo {author} {\bibfnamefont {P.}~\bibnamefont {Yan}}, \bibinfo
  {author} {\bibfnamefont {H.}~\bibnamefont {Huebl}}, \bibinfo {author}
  {\bibfnamefont {S.~T.~B.}\ \bibnamefont {Goennenwein}}, \ and\ \bibinfo
  {author} {\bibfnamefont {G.~E.~W.}\ \bibnamefont {Bauer}},\ }\href {\doibase
  10.1103/PhysRevB.91.094423} {\bibfield  {journal} {\bibinfo  {journal} {Phys.
  Rev. B}\ }\textbf {\bibinfo {volume} {91}},\ \bibinfo {pages} {094423}
  (\bibinfo {year} {2015})}\BibitemShut {NoStop}%
\bibitem [{\citenamefont {Zare~Rameshti}\ \emph {et~al.}(2015)\citenamefont
  {Zare~Rameshti}, \citenamefont {Cao},\ and\ \citenamefont
  {Bauer}}]{BabakPol}%
  \BibitemOpen
  \bibfield  {author} {\bibinfo {author} {\bibfnamefont {B.}~\bibnamefont
  {Zare~Rameshti}}, \bibinfo {author} {\bibfnamefont {Y.}~\bibnamefont {Cao}},
  \ and\ \bibinfo {author} {\bibfnamefont {G.~E.~W.}\ \bibnamefont {Bauer}},\
  }\href {\doibase 10.1103/PhysRevB.91.214430} {\bibfield  {journal} {\bibinfo
  {journal} {Phys. Rev. B}\ }\textbf {\bibinfo {volume} {91}},\ \bibinfo
  {pages} {214430} (\bibinfo {year} {2015})}\BibitemShut {NoStop}%
\bibitem [{\citenamefont {Zare~Rameshti}\ and\ \citenamefont
  {Bauer}(2018)}]{Babak18}%
  \BibitemOpen
  \bibfield  {author} {\bibinfo {author} {\bibfnamefont {B.}~\bibnamefont
  {Zare~Rameshti}}\ and\ \bibinfo {author} {\bibfnamefont {G.~E.~W.}\
  \bibnamefont {Bauer}},\ }\href {\doibase 10.1103/PhysRevB.97.014419}
  {\bibfield  {journal} {\bibinfo  {journal} {Phys. Rev. B}\ }\textbf {\bibinfo
  {volume} {97}},\ \bibinfo {pages} {014419} (\bibinfo {year}
  {2018})}\BibitemShut {NoStop}%
\bibitem [{\citenamefont {McKenzie-Sell}\ \emph {et~al.}(2019)\citenamefont
  {McKenzie-Sell}, \citenamefont {Xie}, \citenamefont {Lee}, \citenamefont
  {Robinson}, \citenamefont {Ciccarelli},\ and\ \citenamefont
  {Haigh}}]{JamesMW}%
  \BibitemOpen
  \bibfield  {author} {\bibinfo {author} {\bibfnamefont {L.}~\bibnamefont
  {McKenzie-Sell}}, \bibinfo {author} {\bibfnamefont {J.}~\bibnamefont {Xie}},
  \bibinfo {author} {\bibfnamefont {C.-M.}\ \bibnamefont {Lee}}, \bibinfo
  {author} {\bibfnamefont {J.~W.~A.}\ \bibnamefont {Robinson}}, \bibinfo
  {author} {\bibfnamefont {C.}~\bibnamefont {Ciccarelli}}, \ and\ \bibinfo
  {author} {\bibfnamefont {J.~A.}\ \bibnamefont {Haigh}},\ }\href {\doibase
  10.1103/PhysRevB.99.140414} {\bibfield  {journal} {\bibinfo  {journal} {Phys.
  Rev. B}\ }\textbf {\bibinfo {volume} {99}},\ \bibinfo {pages} {140414}
  (\bibinfo {year} {2019})}\BibitemShut {NoStop}%
\bibitem [{\citenamefont {Kajiwara}\ \emph {et~al.}(2010)\citenamefont
  {Kajiwara}, \citenamefont {Harii}, \citenamefont {Takahashi}, \citenamefont
  {Ohe}, \citenamefont {Uchida}, \citenamefont {Mizuguchi}, \citenamefont
  {Umezawa}, \citenamefont {Kawai}, \citenamefont {Ando}, \citenamefont
  {Takanashi}, \citenamefont {Maekawa},\ and\ \citenamefont
  {Saitoh}}]{Kajiwara10}%
  \BibitemOpen
  \bibfield  {author} {\bibinfo {author} {\bibfnamefont {Y.}~\bibnamefont
  {Kajiwara}}, \bibinfo {author} {\bibfnamefont {K.}~\bibnamefont {Harii}},
  \bibinfo {author} {\bibfnamefont {S.}~\bibnamefont {Takahashi}}, \bibinfo
  {author} {\bibfnamefont {J.}~\bibnamefont {Ohe}}, \bibinfo {author}
  {\bibfnamefont {K.}~\bibnamefont {Uchida}}, \bibinfo {author} {\bibfnamefont
  {M.}~\bibnamefont {Mizuguchi}}, \bibinfo {author} {\bibfnamefont
  {H.}~\bibnamefont {Umezawa}}, \bibinfo {author} {\bibfnamefont
  {H.}~\bibnamefont {Kawai}}, \bibinfo {author} {\bibfnamefont
  {K.}~\bibnamefont {Ando}}, \bibinfo {author} {\bibfnamefont {K.}~\bibnamefont
  {Takanashi}}, \bibinfo {author} {\bibfnamefont {S.}~\bibnamefont {Maekawa}},
  \ and\ \bibinfo {author} {\bibfnamefont {E.}~\bibnamefont {Saitoh}},\ }\href
  {\doibase 10.1038/nature08876} {\bibfield  {journal} {\bibinfo  {journal}
  {Nature}\ }\textbf {\bibinfo {volume} {464}},\ \bibinfo {pages} {262}
  (\bibinfo {year} {2010})}\BibitemShut {NoStop}%
\bibitem [{\citenamefont {Goennenwein}\ \emph {et~al.}(2015)\citenamefont
  {Goennenwein}, \citenamefont {Schlitz}, \citenamefont {Pernpeintner},
  \citenamefont {Ganzhorn}, \citenamefont {Althammer}, \citenamefont {Gross},\
  and\ \citenamefont {Huebl}}]{SMR_NL1}%
  \BibitemOpen
  \bibfield  {author} {\bibinfo {author} {\bibfnamefont {S.~T.~B.}\
  \bibnamefont {Goennenwein}}, \bibinfo {author} {\bibfnamefont
  {R.}~\bibnamefont {Schlitz}}, \bibinfo {author} {\bibfnamefont
  {M.}~\bibnamefont {Pernpeintner}}, \bibinfo {author} {\bibfnamefont
  {K.}~\bibnamefont {Ganzhorn}}, \bibinfo {author} {\bibfnamefont
  {M.}~\bibnamefont {Althammer}}, \bibinfo {author} {\bibfnamefont
  {R.}~\bibnamefont {Gross}}, \ and\ \bibinfo {author} {\bibfnamefont
  {H.}~\bibnamefont {Huebl}},\ }\href {\doibase 10.1063/1.4935074} {\bibfield
  {journal} {\bibinfo  {journal} {Applied Physics Letters}\ }\textbf {\bibinfo
  {volume} {107}},\ \bibinfo {pages} {172405} (\bibinfo {year} {2015})},\
  \Eprint {http://arxiv.org/abs/https://doi.org/10.1063/1.4935074}
  {https://doi.org/10.1063/1.4935074} \BibitemShut {NoStop}%
\bibitem [{\citenamefont {Zhang}\ \emph {et~al.}(2016)\citenamefont {Zhang},
  \citenamefont {Zhu}, \citenamefont {Zou},\ and\ \citenamefont
  {Tang}}]{ZhangWGM16}%
  \BibitemOpen
  \bibfield  {author} {\bibinfo {author} {\bibfnamefont {X.}~\bibnamefont
  {Zhang}}, \bibinfo {author} {\bibfnamefont {N.}~\bibnamefont {Zhu}}, \bibinfo
  {author} {\bibfnamefont {C.-L.}\ \bibnamefont {Zou}}, \ and\ \bibinfo
  {author} {\bibfnamefont {H.~X.}\ \bibnamefont {Tang}},\ }\href@noop {}
  {\bibfield  {journal} {\bibinfo  {journal} {Phys. Rev. Lett.}\ }\textbf
  {\bibinfo {volume} {117}},\ \bibinfo {pages} {123605} (\bibinfo {year}
  {2016})}\BibitemShut {NoStop}%
\bibitem [{\citenamefont {Osada}\ \emph {et~al.}(2016)\citenamefont {Osada},
  \citenamefont {Hisatomi}, \citenamefont {Noguchi}, \citenamefont {Tabuchi},
  \citenamefont {Yamazaki}, \citenamefont {Usami}, \citenamefont {Sadgrove},
  \citenamefont {Yalla}, \citenamefont {Nomura},\ and\ \citenamefont
  {Nakamura}}]{Osada16}%
  \BibitemOpen
  \bibfield  {author} {\bibinfo {author} {\bibfnamefont {A.}~\bibnamefont
  {Osada}}, \bibinfo {author} {\bibfnamefont {R.}~\bibnamefont {Hisatomi}},
  \bibinfo {author} {\bibfnamefont {A.}~\bibnamefont {Noguchi}}, \bibinfo
  {author} {\bibfnamefont {Y.}~\bibnamefont {Tabuchi}}, \bibinfo {author}
  {\bibfnamefont {R.}~\bibnamefont {Yamazaki}}, \bibinfo {author}
  {\bibfnamefont {K.}~\bibnamefont {Usami}}, \bibinfo {author} {\bibfnamefont
  {M.}~\bibnamefont {Sadgrove}}, \bibinfo {author} {\bibfnamefont
  {R.}~\bibnamefont {Yalla}}, \bibinfo {author} {\bibfnamefont
  {M.}~\bibnamefont {Nomura}}, \ and\ \bibinfo {author} {\bibfnamefont
  {Y.}~\bibnamefont {Nakamura}},\ }\href@noop {} {\bibfield  {journal}
  {\bibinfo  {journal} {Phys. Rev. Lett.}\ }\textbf {\bibinfo {volume} {116}},\
  \bibinfo {pages} {223601} (\bibinfo {year} {2016})}\BibitemShut {NoStop}%
\bibitem [{\citenamefont {Haigh}\ \emph {et~al.}(2016)\citenamefont {Haigh},
  \citenamefont {Nunnenkamp}, \citenamefont {Ramsay},\ and\ \citenamefont
  {Ferguson}}]{JamesWGM}%
  \BibitemOpen
  \bibfield  {author} {\bibinfo {author} {\bibfnamefont {J.~A.}\ \bibnamefont
  {Haigh}}, \bibinfo {author} {\bibfnamefont {A.}~\bibnamefont {Nunnenkamp}},
  \bibinfo {author} {\bibfnamefont {A.~J.}\ \bibnamefont {Ramsay}}, \ and\
  \bibinfo {author} {\bibfnamefont {A.~J.}\ \bibnamefont {Ferguson}},\
  }\href@noop {} {\bibfield  {journal} {\bibinfo  {journal} {Phys. Rev. Lett.}\
  }\textbf {\bibinfo {volume} {117}},\ \bibinfo {pages} {133602} (\bibinfo
  {year} {2016})}\BibitemShut {NoStop}%
\bibitem [{\citenamefont {Wood}\ and\ \citenamefont
  {Remeika}(1967)}]{WoodRemeika}%
  \BibitemOpen
  \bibfield  {author} {\bibinfo {author} {\bibfnamefont {D.~L.}\ \bibnamefont
  {Wood}}\ and\ \bibinfo {author} {\bibfnamefont {J.~P.}\ \bibnamefont
  {Remeika}},\ }\href {\doibase 10.1063/1.1709476} {\bibfield  {journal}
  {\bibinfo  {journal} {Journal of Applied Physics}\ }\textbf {\bibinfo
  {volume} {38}},\ \bibinfo {pages} {1038} (\bibinfo {year} {1967})},\ \Eprint
  {http://arxiv.org/abs/https://doi.org/10.1063/1.1709476}
  {https://doi.org/10.1063/1.1709476} \BibitemShut {NoStop}%
\bibitem [{\citenamefont {Lacklison}\ \emph
  {et~al.}(1973{\natexlab{a}})\citenamefont {Lacklison}, \citenamefont {Scott},
  \citenamefont {Ralph},\ and\ \citenamefont {Page}}]{Lacklison}%
  \BibitemOpen
  \bibfield  {author} {\bibinfo {author} {\bibfnamefont {D.}~\bibnamefont
  {Lacklison}}, \bibinfo {author} {\bibfnamefont {G.}~\bibnamefont {Scott}},
  \bibinfo {author} {\bibfnamefont {H.}~\bibnamefont {Ralph}}, \ and\ \bibinfo
  {author} {\bibfnamefont {J.}~\bibnamefont {Page}},\ }\href {\doibase
  10.1109/TMAG.1973.1067670} {\bibfield  {journal} {\bibinfo  {journal} {IEEE
  Transactions on Magnetics}\ }\textbf {\bibinfo {volume} {9}},\ \bibinfo
  {pages} {457} (\bibinfo {year} {1973}{\natexlab{a}})}\BibitemShut {NoStop}%
\bibitem [{\citenamefont {Haigh}\ \emph {et~al.}(2015)\citenamefont {Haigh},
  \citenamefont {Langenfeld}, \citenamefont {Lambert}, \citenamefont
  {Baumberg}, \citenamefont {Ramsay}, \citenamefont {Nunnenkamp},\ and\
  \citenamefont {Ferguson}}]{James15}%
  \BibitemOpen
  \bibfield  {author} {\bibinfo {author} {\bibfnamefont {J.~A.}\ \bibnamefont
  {Haigh}}, \bibinfo {author} {\bibfnamefont {S.}~\bibnamefont {Langenfeld}},
  \bibinfo {author} {\bibfnamefont {N.~J.}\ \bibnamefont {Lambert}}, \bibinfo
  {author} {\bibfnamefont {J.~J.}\ \bibnamefont {Baumberg}}, \bibinfo {author}
  {\bibfnamefont {A.~J.}\ \bibnamefont {Ramsay}}, \bibinfo {author}
  {\bibfnamefont {A.}~\bibnamefont {Nunnenkamp}}, \ and\ \bibinfo {author}
  {\bibfnamefont {A.~J.}\ \bibnamefont {Ferguson}},\ }\href@noop {} {\bibfield
  {journal} {\bibinfo  {journal} {Phys. Rev. A}\ }\textbf {\bibinfo {volume}
  {92}},\ \bibinfo {pages} {063845} (\bibinfo {year} {2015})}\BibitemShut
  {NoStop}%
\bibitem [{\citenamefont {Wettling}\ \emph {et~al.}(1975)\citenamefont
  {Wettling}, \citenamefont {Cottam},\ and\ \citenamefont
  {Sandercock}}]{Wettling75}%
  \BibitemOpen
  \bibfield  {author} {\bibinfo {author} {\bibfnamefont {W.}~\bibnamefont
  {Wettling}}, \bibinfo {author} {\bibfnamefont {M.~G.}\ \bibnamefont
  {Cottam}}, \ and\ \bibinfo {author} {\bibfnamefont {J.~R.}\ \bibnamefont
  {Sandercock}},\ }\href {http://stacks.iop.org/0022-3719/8/i=2/a=014}
  {\bibfield  {journal} {\bibinfo  {journal} {Journal of Physics C: Solid State
  Physics}\ }\textbf {\bibinfo {volume} {8}},\ \bibinfo {pages} {211} (\bibinfo
  {year} {1975})}\BibitemShut {NoStop}%
\bibitem [{\citenamefont {Borovik-Romanov}\ and\ \citenamefont
  {Kreines}(1982)}]{Borovik82}%
  \BibitemOpen
  \bibfield  {author} {\bibinfo {author} {\bibfnamefont {A.}~\bibnamefont
  {Borovik-Romanov}}\ and\ \bibinfo {author} {\bibfnamefont {N.}~\bibnamefont
  {Kreines}},\ }\href {\doibase http://dx.doi.org/10.1016/0370-1573(82)90118-1}
  {\bibfield  {journal} {\bibinfo  {journal} {Physics Reports}\ }\textbf
  {\bibinfo {volume} {81}},\ \bibinfo {pages} {351 } (\bibinfo {year}
  {1982})}\BibitemShut {NoStop}%
\bibitem [{\citenamefont {Sebastian}\ \emph {et~al.}(2015)\citenamefont
  {Sebastian}, \citenamefont {Schultheiss}, \citenamefont {Obry}, \citenamefont
  {Hillebrands}, \citenamefont {Schultheiss},\ and\ \citenamefont
  {Obry}}]{Sebastian15Rev}%
  \BibitemOpen
  \bibfield  {author} {\bibinfo {author} {\bibfnamefont {T.}~\bibnamefont
  {Sebastian}}, \bibinfo {author} {\bibfnamefont {K.}~\bibnamefont
  {Schultheiss}}, \bibinfo {author} {\bibfnamefont {B.}~\bibnamefont {Obry}},
  \bibinfo {author} {\bibfnamefont {B.}~\bibnamefont {Hillebrands}}, \bibinfo
  {author} {\bibfnamefont {H.}~\bibnamefont {Schultheiss}}, \ and\ \bibinfo
  {author} {\bibfnamefont {B.}~\bibnamefont {Obry}},\ }\href {\doibase
  10.3389/fphy.2015.00035} {\bibfield  {journal} {\bibinfo  {journal}
  {Frontiers in Physics}\ }\textbf {\bibinfo {volume} {3}} (\bibinfo {year}
  {2015}),\ 10.3389/fphy.2015.00035}\BibitemShut {NoStop}%
\bibitem [{\citenamefont {Hisatomi}\ \emph {et~al.}(2016)\citenamefont
  {Hisatomi}, \citenamefont {Osada}, \citenamefont {Tabuchi}, \citenamefont
  {Ishikawa}, \citenamefont {Noguchi}, \citenamefont {Yamazaki}, \citenamefont
  {Usami},\ and\ \citenamefont {Nakamura}}]{Hisatomi16}%
  \BibitemOpen
  \bibfield  {author} {\bibinfo {author} {\bibfnamefont {R.}~\bibnamefont
  {Hisatomi}}, \bibinfo {author} {\bibfnamefont {A.}~\bibnamefont {Osada}},
  \bibinfo {author} {\bibfnamefont {Y.}~\bibnamefont {Tabuchi}}, \bibinfo
  {author} {\bibfnamefont {T.}~\bibnamefont {Ishikawa}}, \bibinfo {author}
  {\bibfnamefont {A.}~\bibnamefont {Noguchi}}, \bibinfo {author} {\bibfnamefont
  {R.}~\bibnamefont {Yamazaki}}, \bibinfo {author} {\bibfnamefont
  {K.}~\bibnamefont {Usami}}, \ and\ \bibinfo {author} {\bibfnamefont
  {Y.}~\bibnamefont {Nakamura}},\ }\href@noop {} {\bibfield  {journal}
  {\bibinfo  {journal} {Phys. Rev. B}\ }\textbf {\bibinfo {volume} {93}},\
  \bibinfo {pages} {174427} (\bibinfo {year} {2016})}\BibitemShut {NoStop}%
\bibitem [{\citenamefont {Osada}\ \emph
  {et~al.}(2018{\natexlab{a}})\citenamefont {Osada}, \citenamefont {Gloppe},
  \citenamefont {Hisatomi}, \citenamefont {Noguchi}, \citenamefont {Yamazaki},
  \citenamefont {Nomura}, \citenamefont {Nakamura},\ and\ \citenamefont
  {Usami}}]{Osada18_Exp}%
  \BibitemOpen
  \bibfield  {author} {\bibinfo {author} {\bibfnamefont {A.}~\bibnamefont
  {Osada}}, \bibinfo {author} {\bibfnamefont {A.}~\bibnamefont {Gloppe}},
  \bibinfo {author} {\bibfnamefont {R.}~\bibnamefont {Hisatomi}}, \bibinfo
  {author} {\bibfnamefont {A.}~\bibnamefont {Noguchi}}, \bibinfo {author}
  {\bibfnamefont {R.}~\bibnamefont {Yamazaki}}, \bibinfo {author}
  {\bibfnamefont {M.}~\bibnamefont {Nomura}}, \bibinfo {author} {\bibfnamefont
  {Y.}~\bibnamefont {Nakamura}}, \ and\ \bibinfo {author} {\bibfnamefont
  {K.}~\bibnamefont {Usami}},\ }\href {\doibase 10.1103/PhysRevLett.120.133602}
  {\bibfield  {journal} {\bibinfo  {journal} {Phys. Rev. Lett.}\ }\textbf
  {\bibinfo {volume} {120}},\ \bibinfo {pages} {133602} (\bibinfo {year}
  {2018}{\natexlab{a}})}\BibitemShut {NoStop}%
\bibitem [{\citenamefont {Haigh}\ \emph {et~al.}(2018)\citenamefont {Haigh},
  \citenamefont {Lambert}, \citenamefont {Sharma}, \citenamefont {Blanter},
  \citenamefont {Bauer},\ and\ \citenamefont {Ramsay}}]{JamesLowL}%
  \BibitemOpen
  \bibfield  {author} {\bibinfo {author} {\bibfnamefont {J.~A.}\ \bibnamefont
  {Haigh}}, \bibinfo {author} {\bibfnamefont {N.~J.}\ \bibnamefont {Lambert}},
  \bibinfo {author} {\bibfnamefont {S.}~\bibnamefont {Sharma}}, \bibinfo
  {author} {\bibfnamefont {Y.~M.}\ \bibnamefont {Blanter}}, \bibinfo {author}
  {\bibfnamefont {G.~E.~W.}\ \bibnamefont {Bauer}}, \ and\ \bibinfo {author}
  {\bibfnamefont {A.~J.}\ \bibnamefont {Ramsay}},\ }\href {\doibase
  10.1103/PhysRevB.97.214423} {\bibfield  {journal} {\bibinfo  {journal} {Phys.
  Rev. B}\ }\textbf {\bibinfo {volume} {97}},\ \bibinfo {pages} {214423}
  (\bibinfo {year} {2018})}\BibitemShut {NoStop}%
\bibitem [{\citenamefont {Pantazopoulos}\ \emph {et~al.}(2017)\citenamefont
  {Pantazopoulos}, \citenamefont {Stefanou}, \citenamefont {Almpanis},\ and\
  \citenamefont {Papanikolaou}}]{Pantazopoulos}%
  \BibitemOpen
  \bibfield  {author} {\bibinfo {author} {\bibfnamefont {P.~A.}\ \bibnamefont
  {Pantazopoulos}}, \bibinfo {author} {\bibfnamefont {N.}~\bibnamefont
  {Stefanou}}, \bibinfo {author} {\bibfnamefont {E.}~\bibnamefont {Almpanis}},
  \ and\ \bibinfo {author} {\bibfnamefont {N.}~\bibnamefont {Papanikolaou}},\
  }\href {\doibase 10.1103/PhysRevB.96.104425} {\bibfield  {journal} {\bibinfo
  {journal} {Phys. Rev. B}\ }\textbf {\bibinfo {volume} {96}},\ \bibinfo
  {pages} {104425} (\bibinfo {year} {2017})}\BibitemShut {NoStop}%
\bibitem [{\citenamefont {{Lachance-Quirion}}\ \emph
  {et~al.}(2019)\citenamefont {{Lachance-Quirion}}, \citenamefont {{Tabuchi}},
  \citenamefont {{Gloppe}}, \citenamefont {{Usami}},\ and\ \citenamefont
  {{Nakamura}}}]{MagPhotonRev}%
  \BibitemOpen
  \bibfield  {author} {\bibinfo {author} {\bibfnamefont {D.}~\bibnamefont
  {{Lachance-Quirion}}}, \bibinfo {author} {\bibfnamefont {Y.}~\bibnamefont
  {{Tabuchi}}}, \bibinfo {author} {\bibfnamefont {A.}~\bibnamefont {{Gloppe}}},
  \bibinfo {author} {\bibfnamefont {K.}~\bibnamefont {{Usami}}}, \ and\
  \bibinfo {author} {\bibfnamefont {Y.}~\bibnamefont {{Nakamura}}},\
  }\href@noop {} {\bibfield  {journal} {\bibinfo  {journal} {ArXiv e-prints}\ }
  (\bibinfo {year} {2019})},\ \Eprint {http://arxiv.org/abs/1902.03024}
  {1902.03024 [quant-ph]} \BibitemShut {NoStop}%
\bibitem [{\citenamefont {Liu}\ \emph {et~al.}(2016)\citenamefont {Liu},
  \citenamefont {Zhang}, \citenamefont {Tang},\ and\ \citenamefont
  {Flatt\'e}}]{Tianyu16}%
  \BibitemOpen
  \bibfield  {author} {\bibinfo {author} {\bibfnamefont {T.}~\bibnamefont
  {Liu}}, \bibinfo {author} {\bibfnamefont {X.}~\bibnamefont {Zhang}}, \bibinfo
  {author} {\bibfnamefont {H.~X.}\ \bibnamefont {Tang}}, \ and\ \bibinfo
  {author} {\bibfnamefont {M.~E.}\ \bibnamefont {Flatt\'e}},\ }\href@noop {}
  {\bibfield  {journal} {\bibinfo  {journal} {Phys. Rev. B}\ }\textbf {\bibinfo
  {volume} {94}},\ \bibinfo {pages} {060405} (\bibinfo {year}
  {2016})}\BibitemShut {NoStop}%
\bibitem [{\citenamefont {Kusminskiy}\ \emph {et~al.}(2016)\citenamefont
  {Kusminskiy}, \citenamefont {Tang},\ and\ \citenamefont
  {Marquardt}}]{Silvia16}%
  \BibitemOpen
  \bibfield  {author} {\bibinfo {author} {\bibfnamefont {S.~V.}\ \bibnamefont
  {Kusminskiy}}, \bibinfo {author} {\bibfnamefont {H.~X.}\ \bibnamefont
  {Tang}}, \ and\ \bibinfo {author} {\bibfnamefont {F.}~\bibnamefont
  {Marquardt}},\ }\href@noop {} {\bibfield  {journal} {\bibinfo  {journal}
  {Phys. Rev. A}\ }\textbf {\bibinfo {volume} {94}},\ \bibinfo {pages} {033821}
  (\bibinfo {year} {2016})}\BibitemShut {NoStop}%
\bibitem [{\citenamefont {Sharma}\ \emph {et~al.}(2018)\citenamefont {Sharma},
  \citenamefont {Blanter},\ and\ \citenamefont {Bauer}}]{OMagCool}%
  \BibitemOpen
  \bibfield  {author} {\bibinfo {author} {\bibfnamefont {S.}~\bibnamefont
  {Sharma}}, \bibinfo {author} {\bibfnamefont {Y.~M.}\ \bibnamefont {Blanter}},
  \ and\ \bibinfo {author} {\bibfnamefont {G.~E.~W.}\ \bibnamefont {Bauer}},\
  }\href {\doibase 10.1103/PhysRevLett.121.087205} {\bibfield  {journal}
  {\bibinfo  {journal} {Phys. Rev. Lett.}\ }\textbf {\bibinfo {volume} {121}},\
  \bibinfo {pages} {087205} (\bibinfo {year} {2018})}\BibitemShut {NoStop}%
\bibitem [{\citenamefont {Sharma}\ \emph {et~al.}(2017)\citenamefont {Sharma},
  \citenamefont {Blanter},\ and\ \citenamefont {Bauer}}]{WGMOptoMag}%
  \BibitemOpen
  \bibfield  {author} {\bibinfo {author} {\bibfnamefont {S.}~\bibnamefont
  {Sharma}}, \bibinfo {author} {\bibfnamefont {Y.~M.}\ \bibnamefont {Blanter}},
  \ and\ \bibinfo {author} {\bibfnamefont {G.~E.~W.}\ \bibnamefont {Bauer}},\
  }\href@noop {} {\bibfield  {journal} {\bibinfo  {journal} {Phys. Rev. B}\
  }\textbf {\bibinfo {volume} {96}},\ \bibinfo {pages} {094412} (\bibinfo
  {year} {2017})}\BibitemShut {NoStop}%
\bibitem [{\citenamefont {Osada}\ \emph
  {et~al.}(2018{\natexlab{b}})\citenamefont {Osada}, \citenamefont {Gloppe},
  \citenamefont {Nakamura},\ and\ \citenamefont {Usami}}]{Osada18_Th}%
  \BibitemOpen
  \bibfield  {author} {\bibinfo {author} {\bibfnamefont {A.}~\bibnamefont
  {Osada}}, \bibinfo {author} {\bibfnamefont {A.}~\bibnamefont {Gloppe}},
  \bibinfo {author} {\bibfnamefont {Y.}~\bibnamefont {Nakamura}}, \ and\
  \bibinfo {author} {\bibfnamefont {K.}~\bibnamefont {Usami}},\ }\href
  {\doibase 10.1088/1367-2630/aae4b1} {\bibfield  {journal} {\bibinfo
  {journal} {New Journal of Physics}\ }\textbf {\bibinfo {volume} {20}},\
  \bibinfo {pages} {103018} (\bibinfo {year} {2018}{\natexlab{b}})}\BibitemShut
  {NoStop}%
\bibitem [{\citenamefont {Almpanis}(2018)}]{Evangelos}%
  \BibitemOpen
  \bibfield  {author} {\bibinfo {author} {\bibfnamefont {E.}~\bibnamefont
  {Almpanis}},\ }\href {\doibase 10.1103/PhysRevB.97.184406} {\bibfield
  {journal} {\bibinfo  {journal} {Phys. Rev. B}\ }\textbf {\bibinfo {volume}
  {97}},\ \bibinfo {pages} {184406} (\bibinfo {year} {2018})}\BibitemShut
  {NoStop}%
\bibitem [{\citenamefont {Berzhansky}\ \emph {et~al.}(2013)\citenamefont
  {Berzhansky}, \citenamefont {Mikhailova}, \citenamefont {Shaposhnikov},
  \citenamefont {Prokopov}, \citenamefont {Karavainikov}, \citenamefont
  {Kotov}, \citenamefont {Balabanov},\ and\ \citenamefont
  {Burkov}}]{Berzhansky13}%
  \BibitemOpen
  \bibfield  {author} {\bibinfo {author} {\bibfnamefont {V.}~\bibnamefont
  {Berzhansky}}, \bibinfo {author} {\bibfnamefont {T.}~\bibnamefont
  {Mikhailova}}, \bibinfo {author} {\bibfnamefont {A.}~\bibnamefont
  {Shaposhnikov}}, \bibinfo {author} {\bibfnamefont {A.}~\bibnamefont
  {Prokopov}}, \bibinfo {author} {\bibfnamefont {A.}~\bibnamefont
  {Karavainikov}}, \bibinfo {author} {\bibfnamefont {V.}~\bibnamefont {Kotov}},
  \bibinfo {author} {\bibfnamefont {D.}~\bibnamefont {Balabanov}}, \ and\
  \bibinfo {author} {\bibfnamefont {V.}~\bibnamefont {Burkov}},\ }\href
  {\doibase 10.1364/AO.52.006599} {\bibfield  {journal} {\bibinfo  {journal}
  {Appl. Opt.}\ }\textbf {\bibinfo {volume} {52}},\ \bibinfo {pages} {6599}
  (\bibinfo {year} {2013})}\BibitemShut {NoStop}%
\bibitem [{\citenamefont {{Chang}}\ \emph {et~al.}(2014)\citenamefont
  {{Chang}}, \citenamefont {{Li}}, \citenamefont {{Zhang}}, \citenamefont
  {{Liu}}, \citenamefont {{Hoffmann}}, \citenamefont {{Deng}},\ and\
  \citenamefont {{Wu}}}]{Chang14}%
  \BibitemOpen
  \bibfield  {author} {\bibinfo {author} {\bibfnamefont {H.}~\bibnamefont
  {{Chang}}}, \bibinfo {author} {\bibfnamefont {P.}~\bibnamefont {{Li}}},
  \bibinfo {author} {\bibfnamefont {W.}~\bibnamefont {{Zhang}}}, \bibinfo
  {author} {\bibfnamefont {T.}~\bibnamefont {{Liu}}}, \bibinfo {author}
  {\bibfnamefont {A.}~\bibnamefont {{Hoffmann}}}, \bibinfo {author}
  {\bibfnamefont {L.}~\bibnamefont {{Deng}}}, \ and\ \bibinfo {author}
  {\bibfnamefont {M.}~\bibnamefont {{Wu}}},\ }\href {\doibase
  10.1109/LMAG.2014.2350958} {\bibfield  {journal} {\bibinfo  {journal} {IEEE
  Magnetics Letters}\ }\textbf {\bibinfo {volume} {5}},\ \bibinfo {pages} {1}
  (\bibinfo {year} {2014})}\BibitemShut {NoStop}%
\bibitem [{\citenamefont {Hauser}\ \emph {et~al.}(2016)\citenamefont {Hauser},
  \citenamefont {Richter}, \citenamefont {Homonnay}, \citenamefont
  {Eisenschmidt}, \citenamefont {Qaid}, \citenamefont {Deniz}, \citenamefont
  {Hesse}, \citenamefont {Sawicki}, \citenamefont {Ebbinghaus},\ and\
  \citenamefont {Schmidt}}]{Hauser16}%
  \BibitemOpen
  \bibfield  {author} {\bibinfo {author} {\bibfnamefont {C.}~\bibnamefont
  {Hauser}}, \bibinfo {author} {\bibfnamefont {T.}~\bibnamefont {Richter}},
  \bibinfo {author} {\bibfnamefont {N.}~\bibnamefont {Homonnay}}, \bibinfo
  {author} {\bibfnamefont {C.}~\bibnamefont {Eisenschmidt}}, \bibinfo {author}
  {\bibfnamefont {M.}~\bibnamefont {Qaid}}, \bibinfo {author} {\bibfnamefont
  {H.}~\bibnamefont {Deniz}}, \bibinfo {author} {\bibfnamefont
  {D.}~\bibnamefont {Hesse}}, \bibinfo {author} {\bibfnamefont
  {M.}~\bibnamefont {Sawicki}}, \bibinfo {author} {\bibfnamefont {S.~G.}\
  \bibnamefont {Ebbinghaus}}, \ and\ \bibinfo {author} {\bibfnamefont
  {G.}~\bibnamefont {Schmidt}},\ }\href {\doibase 10.1038/srep20827} {\bibfield
   {journal} {\bibinfo  {journal} {Scientific Reports}\ }\textbf {\bibinfo
  {volume} {6}},\ \bibinfo {pages} {20827} (\bibinfo {year} {2016})},\ \bibinfo
  {note} {article}\BibitemShut {NoStop}%
\bibitem [{\citenamefont {Graf}\ \emph {et~al.}(2018)\citenamefont {Graf},
  \citenamefont {Pfeifer}, \citenamefont {Marquardt},\ and\ \citenamefont
  {Viola~Kusminskiy}}]{Graf_Vortex}%
  \BibitemOpen
  \bibfield  {author} {\bibinfo {author} {\bibfnamefont {J.}~\bibnamefont
  {Graf}}, \bibinfo {author} {\bibfnamefont {H.}~\bibnamefont {Pfeifer}},
  \bibinfo {author} {\bibfnamefont {F.}~\bibnamefont {Marquardt}}, \ and\
  \bibinfo {author} {\bibfnamefont {S.}~\bibnamefont {Viola~Kusminskiy}},\
  }\href {\doibase 10.1103/PhysRevB.98.241406} {\bibfield  {journal} {\bibinfo
  {journal} {Phys. Rev. B}\ }\textbf {\bibinfo {volume} {98}},\ \bibinfo
  {pages} {241406} (\bibinfo {year} {2018})}\BibitemShut {NoStop}%
\bibitem [{\citenamefont {Stancil}\ and\ \citenamefont
  {Prabhakar}(2009)}]{StanPrabh}%
  \BibitemOpen
  \bibfield  {author} {\bibinfo {author} {\bibfnamefont {D.~D.}\ \bibnamefont
  {Stancil}}\ and\ \bibinfo {author} {\bibfnamefont {A.}~\bibnamefont
  {Prabhakar}},\ }\href@noop {} {\emph {\bibinfo {title} {Spin Waves: Theory
  and Applications}}}\ (\bibinfo  {publisher} {Springer US},\ \bibinfo {year}
  {2009})\BibitemShut {NoStop}%
\bibitem [{\citenamefont {Kalinikos}\ and\ \citenamefont
  {Slavin}(1986)}]{KalinikosSlavin}%
  \BibitemOpen
  \bibfield  {author} {\bibinfo {author} {\bibfnamefont {B.~A.}\ \bibnamefont
  {Kalinikos}}\ and\ \bibinfo {author} {\bibfnamefont {A.~N.}\ \bibnamefont
  {Slavin}},\ }\href {\doibase 10.1088/0022-3719/19/35/014} {\bibfield
  {journal} {\bibinfo  {journal} {Journal of Physics C: Solid State Physics}\
  }\textbf {\bibinfo {volume} {19}},\ \bibinfo {pages} {7013} (\bibinfo {year}
  {1986})}\BibitemShut {NoStop}%
\bibitem [{\citenamefont {Rych{\l}y}\ \emph {et~al.}(2018)\citenamefont
  {Rych{\l}y}, \citenamefont {Tkachenko}, \citenamefont {K{\l}os},
  \citenamefont {Kuchko},\ and\ \citenamefont {Krawczyk}}]{CylinderMagnons18}%
  \BibitemOpen
  \bibfield  {author} {\bibinfo {author} {\bibfnamefont {J.}~\bibnamefont
  {Rych{\l}y}}, \bibinfo {author} {\bibfnamefont {V.~S.}\ \bibnamefont
  {Tkachenko}}, \bibinfo {author} {\bibfnamefont {J.~W.}\ \bibnamefont
  {K{\l}os}}, \bibinfo {author} {\bibfnamefont {A.}~\bibnamefont {Kuchko}}, \
  and\ \bibinfo {author} {\bibfnamefont {M.}~\bibnamefont {Krawczyk}},\ }\href
  {\doibase 10.1088/1361-6463/aaf2fc} {\bibfield  {journal} {\bibinfo
  {journal} {Journal of Physics D: Applied Physics}\ }\textbf {\bibinfo
  {volume} {52}},\ \bibinfo {pages} {075003} (\bibinfo {year}
  {2018})}\BibitemShut {NoStop}%
\bibitem [{\citenamefont {Wolfram}\ and\ \citenamefont
  {De~Wames}(1970)}]{WamesWolfram_Linewidth}%
  \BibitemOpen
  \bibfield  {author} {\bibinfo {author} {\bibfnamefont {T.}~\bibnamefont
  {Wolfram}}\ and\ \bibinfo {author} {\bibfnamefont {R.~E.}\ \bibnamefont
  {De~Wames}},\ }\href {\doibase 10.1103/PhysRevB.1.4358} {\bibfield  {journal}
  {\bibinfo  {journal} {Phys. Rev. B}\ }\textbf {\bibinfo {volume} {1}},\
  \bibinfo {pages} {4358} (\bibinfo {year} {1970})}\BibitemShut {NoStop}%
\bibitem [{\citenamefont {Camley}\ and\ \citenamefont
  {Mills}(1978)}]{CamleyMills}%
  \BibitemOpen
  \bibfield  {author} {\bibinfo {author} {\bibfnamefont {R.~E.}\ \bibnamefont
  {Camley}}\ and\ \bibinfo {author} {\bibfnamefont {D.~L.}\ \bibnamefont
  {Mills}},\ }\href {\doibase 10.1103/PhysRevB.18.4821} {\bibfield  {journal}
  {\bibinfo  {journal} {Phys. Rev. B}\ }\textbf {\bibinfo {volume} {18}},\
  \bibinfo {pages} {4821} (\bibinfo {year} {1978})}\BibitemShut {NoStop}%
\bibitem [{\citenamefont {Borghese}\ and\ \citenamefont
  {De~Gasperis}(1980)}]{CamilloPaolo}%
  \BibitemOpen
  \bibfield  {author} {\bibinfo {author} {\bibfnamefont {C.}~\bibnamefont
  {Borghese}}\ and\ \bibinfo {author} {\bibfnamefont {P.}~\bibnamefont
  {De~Gasperis}},\ }\href {\doibase 10.1063/1.327497} {\bibfield  {journal}
  {\bibinfo  {journal} {Journal of Applied Physics}\ }\textbf {\bibinfo
  {volume} {51}},\ \bibinfo {pages} {5425} (\bibinfo {year} {1980})},\ \Eprint
  {http://arxiv.org/abs/https://aip.scitation.org/doi/pdf/10.1063/1.327497}
  {https://aip.scitation.org/doi/pdf/10.1063/1.327497} \BibitemShut {NoStop}%
\bibitem [{\citenamefont {Wames}\ and\ \citenamefont
  {Wolfram}(1970)}]{WamesWolfram_Detailed}%
  \BibitemOpen
  \bibfield  {author} {\bibinfo {author} {\bibfnamefont {R.~E.~D.}\
  \bibnamefont {Wames}}\ and\ \bibinfo {author} {\bibfnamefont
  {T.}~\bibnamefont {Wolfram}},\ }\href@noop {} {\bibfield  {journal} {\bibinfo
   {journal} {Journal of Applied Physics}\ }\textbf {\bibinfo {volume} {41}},\
  \bibinfo {pages} {987} (\bibinfo {year} {1970})}\BibitemShut {NoStop}%
\bibitem [{\citenamefont {Oraevsky}(2002)}]{Oraevsky_WGM}%
  \BibitemOpen
  \bibfield  {author} {\bibinfo {author} {\bibfnamefont {A.~N.}\ \bibnamefont
  {Oraevsky}},\ }\href {http://stacks.iop.org/1063-7818/32/i=5/a=R01}
  {\bibfield  {journal} {\bibinfo  {journal} {Quantum Electronics}\ }\textbf
  {\bibinfo {volume} {32}},\ \bibinfo {pages} {377} (\bibinfo {year}
  {2002})}\BibitemShut {NoStop}%
\bibitem [{\citenamefont {Abramowitz}\ and\ \citenamefont
  {Stegun}(1964)}]{AbrSteg}%
  \BibitemOpen
  \bibfield  {author} {\bibinfo {author} {\bibfnamefont {M.}~\bibnamefont
  {Abramowitz}}\ and\ \bibinfo {author} {\bibfnamefont {I.~A.}\ \bibnamefont
  {Stegun}},\ }\href {http://people.math.sfu.ca/~cbm/aands/} {\emph {\bibinfo
  {title} {Handbook of Mathematical Functions}}},\ Applied Mathematics Series\
  (\bibinfo  {publisher} {National Bureau of Standards},\ \bibinfo {year}
  {1964})\BibitemShut {NoStop}%
\bibitem [{\citenamefont {Walker}(1957)}]{WalkerOrig}%
  \BibitemOpen
  \bibfield  {author} {\bibinfo {author} {\bibfnamefont {L.~R.}\ \bibnamefont
  {Walker}},\ }\href {\doibase 10.1103/PhysRev.105.390} {\bibfield  {journal}
  {\bibinfo  {journal} {Phys. Rev.}\ }\textbf {\bibinfo {volume} {105}},\
  \bibinfo {pages} {390} (\bibinfo {year} {1957})}\BibitemShut {NoStop}%
\bibitem [{\citenamefont {Sandercock}\ and\ \citenamefont
  {Wettling}(1973)}]{Sandercock73}%
  \BibitemOpen
  \bibfield  {author} {\bibinfo {author} {\bibfnamefont {J.}~\bibnamefont
  {Sandercock}}\ and\ \bibinfo {author} {\bibfnamefont {W.}~\bibnamefont
  {Wettling}},\ }\href {\doibase
  http://dx.doi.org/10.1016/0038-1098(73)90276-7} {\bibfield  {journal}
  {\bibinfo  {journal} {Solid State Communications}\ }\textbf {\bibinfo
  {volume} {13}},\ \bibinfo {pages} {1729 } (\bibinfo {year}
  {1973})}\BibitemShut {NoStop}%
\bibitem [{\citenamefont {Klingler}\ \emph {et~al.}(2015)\citenamefont
  {Klingler}, \citenamefont {Chumak}, \citenamefont {Mewes}, \citenamefont
  {Khodadadi}, \citenamefont {Mewes}, \citenamefont {Dubs}, \citenamefont
  {Surzhenko}, \citenamefont {Hillebrands},\ and\ \citenamefont
  {Conca}}]{KlinglerDex}%
  \BibitemOpen
  \bibfield  {author} {\bibinfo {author} {\bibfnamefont {S.}~\bibnamefont
  {Klingler}}, \bibinfo {author} {\bibfnamefont {A.}~\bibnamefont {Chumak}},
  \bibinfo {author} {\bibfnamefont {T.}~\bibnamefont {Mewes}}, \bibinfo
  {author} {\bibfnamefont {B.}~\bibnamefont {Khodadadi}}, \bibinfo {author}
  {\bibfnamefont {C.}~\bibnamefont {Mewes}}, \bibinfo {author} {\bibfnamefont
  {C.}~\bibnamefont {Dubs}}, \bibinfo {author} {\bibfnamefont {O.}~\bibnamefont
  {Surzhenko}}, \bibinfo {author} {\bibfnamefont {B.}~\bibnamefont
  {Hillebrands}}, \ and\ \bibinfo {author} {\bibfnamefont {A.}~\bibnamefont
  {Conca}},\ }\href {http://stacks.iop.org/0022-3727/48/i=1/a=015001}
  {\bibfield  {journal} {\bibinfo  {journal} {Journal of Physics D: Applied
  Physics}\ }\textbf {\bibinfo {volume} {48}},\ \bibinfo {pages} {015001}
  (\bibinfo {year} {2015})}\BibitemShut {NoStop}%
\bibitem [{\citenamefont {Deeter}\ \emph {et~al.}(1990)\citenamefont {Deeter},
  \citenamefont {Rose},\ and\ \citenamefont {Day}}]{Deeter90}%
  \BibitemOpen
  \bibfield  {author} {\bibinfo {author} {\bibfnamefont {M.~N.}\ \bibnamefont
  {Deeter}}, \bibinfo {author} {\bibfnamefont {A.~H.}\ \bibnamefont {Rose}}, \
  and\ \bibinfo {author} {\bibfnamefont {G.~W.}\ \bibnamefont {Day}},\ }\href
  {\doibase 10.1109/50.62880} {\bibfield  {journal} {\bibinfo  {journal}
  {Journal of Lightwave Technology}\ }\textbf {\bibinfo {volume} {8}},\
  \bibinfo {pages} {1838} (\bibinfo {year} {1990})}\BibitemShut {NoStop}%
\bibitem [{\citenamefont {Scott}\ and\ \citenamefont
  {Lacklison}(1976)}]{ScottLacklison}%
  \BibitemOpen
  \bibfield  {author} {\bibinfo {author} {\bibfnamefont {G.}~\bibnamefont
  {Scott}}\ and\ \bibinfo {author} {\bibfnamefont {D.}~\bibnamefont
  {Lacklison}},\ }\href {\doibase 10.1109/TMAG.1976.1059031} {\bibfield
  {journal} {\bibinfo  {journal} {IEEE Transactions on Magnetics}\ }\textbf
  {\bibinfo {volume} {12}},\ \bibinfo {pages} {292} (\bibinfo {year}
  {1976})}\BibitemShut {NoStop}%
\bibitem [{\citenamefont {Castera}\ and\ \citenamefont
  {Hepner}(1977)}]{Castera77}%
  \BibitemOpen
  \bibfield  {author} {\bibinfo {author} {\bibfnamefont {J.}~\bibnamefont
  {Castera}}\ and\ \bibinfo {author} {\bibfnamefont {G.}~\bibnamefont
  {Hepner}},\ }\href {\doibase 10.1109/TMAG.1977.1059641} {\bibfield  {journal}
  {\bibinfo  {journal} {IEEE Transactions on Magnetics}\ }\textbf {\bibinfo
  {volume} {13}},\ \bibinfo {pages} {1583} (\bibinfo {year}
  {1977})}\BibitemShut {NoStop}%
\bibitem [{\citenamefont {Kamada}\ and\ \citenamefont
  {Higuchi}(2001)}]{Kamada_2001}%
  \BibitemOpen
  \bibfield  {author} {\bibinfo {author} {\bibfnamefont {O.}~\bibnamefont
  {Kamada}}\ and\ \bibinfo {author} {\bibfnamefont {S.}~\bibnamefont
  {Higuchi}},\ }\href {\doibase 10.1109/20.951038} {\bibfield  {journal}
  {\bibinfo  {journal} {IEEE Transactions on Magnetics}\ }\textbf {\bibinfo
  {volume} {37}},\ \bibinfo {pages} {2013} (\bibinfo {year}
  {2001})}\BibitemShut {NoStop}%
\bibitem [{\citenamefont {Hurben}\ and\ \citenamefont
  {Patton}(1995)}]{Hurben95}%
  \BibitemOpen
  \bibfield  {author} {\bibinfo {author} {\bibfnamefont {M.}~\bibnamefont
  {Hurben}}\ and\ \bibinfo {author} {\bibfnamefont {C.}~\bibnamefont
  {Patton}},\ }\href {\doibase http://dx.doi.org/10.1016/0304-8853(95)90006-3}
  {\bibfield  {journal} {\bibinfo  {journal} {Journal of Magnetism and Magnetic
  Materials}\ }\textbf {\bibinfo {volume} {139}},\ \bibinfo {pages} {263 }
  (\bibinfo {year} {1995})}\BibitemShut {NoStop}%
\bibitem [{\citenamefont {Soohoo}(1965)}]{SoohooBook}%
  \BibitemOpen
  \bibfield  {author} {\bibinfo {author} {\bibfnamefont {R.}~\bibnamefont
  {Soohoo}},\ }\href {https://books.google.nl/books?id=NefvAAAAMAAJ} {\emph
  {\bibinfo {title} {Magnetic Thin Films}}},\ Harper's physics series\
  (\bibinfo  {publisher} {Harper and Row},\ \bibinfo {year} {1965})\BibitemShut
  {NoStop}%
\bibitem [{\citenamefont {Guslienko}\ and\ \citenamefont
  {Slavin}(2005)}]{Guslienko_BD}%
  \BibitemOpen
  \bibfield  {author} {\bibinfo {author} {\bibfnamefont {K.~Y.}\ \bibnamefont
  {Guslienko}}\ and\ \bibinfo {author} {\bibfnamefont {A.~N.}\ \bibnamefont
  {Slavin}},\ }\href {\doibase 10.1103/PhysRevB.72.014463} {\bibfield
  {journal} {\bibinfo  {journal} {Phys. Rev. B}\ }\textbf {\bibinfo {volume}
  {72}},\ \bibinfo {pages} {014463} (\bibinfo {year} {2005})}\BibitemShut
  {NoStop}%
\bibitem [{\citenamefont {Damon}\ and\ \citenamefont
  {Eshbach}(1961)}]{DamEshSlab}%
  \BibitemOpen
  \bibfield  {author} {\bibinfo {author} {\bibfnamefont {R.}~\bibnamefont
  {Damon}}\ and\ \bibinfo {author} {\bibfnamefont {J.}~\bibnamefont
  {Eshbach}},\ }\href {\doibase http://dx.doi.org/10.1016/0022-3697(61)90041-5}
  {\bibfield  {journal} {\bibinfo  {journal} {Journal of Physics and Chemistry
  of Solids}\ }\textbf {\bibinfo {volume} {19}},\ \bibinfo {pages} {308 }
  (\bibinfo {year} {1961})}\BibitemShut {NoStop}%
\bibitem [{\citenamefont {Eshbach}\ and\ \citenamefont
  {Damon}(1960)}]{DamEshSurface}%
  \BibitemOpen
  \bibfield  {author} {\bibinfo {author} {\bibfnamefont {J.~R.}\ \bibnamefont
  {Eshbach}}\ and\ \bibinfo {author} {\bibfnamefont {R.~W.}\ \bibnamefont
  {Damon}},\ }\href {\doibase 10.1103/PhysRev.118.1208} {\bibfield  {journal}
  {\bibinfo  {journal} {Phys. Rev.}\ }\textbf {\bibinfo {volume} {118}},\
  \bibinfo {pages} {1208} (\bibinfo {year} {1960})}\BibitemShut {NoStop}%
\bibitem [{\citenamefont {Lacklison}\ \emph
  {et~al.}(1973{\natexlab{b}})\citenamefont {Lacklison}, \citenamefont {Scott},
  \citenamefont {Ralph},\ and\ \citenamefont {Page}}]{LacklisonBiYIG}%
  \BibitemOpen
  \bibfield  {author} {\bibinfo {author} {\bibfnamefont {D.}~\bibnamefont
  {Lacklison}}, \bibinfo {author} {\bibfnamefont {G.}~\bibnamefont {Scott}},
  \bibinfo {author} {\bibfnamefont {H.}~\bibnamefont {Ralph}}, \ and\ \bibinfo
  {author} {\bibfnamefont {J.}~\bibnamefont {Page}},\ }\href@noop {} {\bibfield
   {journal} {\bibinfo  {journal} {IEEE Transactions on Magnetics}\ }\textbf
  {\bibinfo {volume} {9}},\ \bibinfo {pages} {457} (\bibinfo {year}
  {1973}{\natexlab{b}})}\BibitemShut {NoStop}%
\bibitem [{\citenamefont {Kostylev}(2013)}]{Kostylev}%
  \BibitemOpen
  \bibfield  {author} {\bibinfo {author} {\bibfnamefont {M.}~\bibnamefont
  {Kostylev}},\ }\href {\doibase 10.1063/1.4789962} {\bibfield  {journal}
  {\bibinfo  {journal} {Journal of Applied Physics}\ }\textbf {\bibinfo
  {volume} {113}},\ \bibinfo {pages} {053907} (\bibinfo {year} {2013})},\
  \Eprint {http://arxiv.org/abs/https://doi.org/10.1063/1.4789962}
  {https://doi.org/10.1063/1.4789962} \BibitemShut {NoStop}%
\bibitem [{\citenamefont {Kamra}\ and\ \citenamefont
  {Belzig}(2016)}]{AkashSqueezing}%
  \BibitemOpen
  \bibfield  {author} {\bibinfo {author} {\bibfnamefont {A.}~\bibnamefont
  {Kamra}}\ and\ \bibinfo {author} {\bibfnamefont {W.}~\bibnamefont {Belzig}},\
  }\href {\doibase 10.1103/PhysRevLett.116.146601} {\bibfield  {journal}
  {\bibinfo  {journal} {Phys. Rev. Lett.}\ }\textbf {\bibinfo {volume} {116}},\
  \bibinfo {pages} {146601} (\bibinfo {year} {2016})}\BibitemShut {NoStop}%
\end{thebibliography}
%

\end{document}